\documentclass[10pt,journal,compsoc]{IEEEtran}

\usepackage{xcolor,soul,framed,caption} 

\colorlet{shadecolor}{yellow}
\usepackage[pdftex]{graphicx}
\graphicspath{{../pdf/}{../jpeg/}}
\DeclareGraphicsExtensions{.pdf,.jpeg,.png}

\usepackage[cmex10]{amsmath}
\usepackage{array}
\usepackage{mdwmath}
\usepackage{mdwtab}
\usepackage{eqparbox}
\usepackage{url}
\usepackage{supertabular}                        \usepackage[colorlinks=true,
    linkcolor=blue,
    urlcolor=blue,
    citecolor=blue]{hyperref}    
\usepackage{amsmath,amssymb}
\usepackage{caption}
\usepackage{subcaption}
\usepackage{float}
\usepackage{booktabs}
\usepackage{multirow}
\usepackage{graphicx}
\usepackage{color}
\usepackage{tabularray}
\usepackage{hhline}
\usepackage{wasysym}
\usepackage[ruled, lined, longend, linesnumbered]{algorithm2e}
\begin{document}
\bstctlcite{IEEEexample:BSTcontrol}
    \title{TF-DDRL: A Transformer-enhanced Distributed DRL Technique for Scheduling IoT Applications in Edge and Cloud Computing Environments}
  \author{Zhiyu Wang, Mohammad Goudarzi, and Rajkumar Buyya

  \thanks{Zhiyu Wang and Rajkumar Buyya are with the Cloud Computing and Distributed Systems (CLOUDS) Laboratory, School of Computing and Information Systems, The University of Melbourne, Australia (e-mail: zhiywang1@student.unimelb.edu.au, rbuyya@unimelb.edu.au).}
  \thanks{Mohammad Goudarzi is with the Faculty of Information Technology, Monash University, Australia (email: mohammad.goudarzi@monash.edu)}
}


\maketitle

\begin{abstract}
With the continuous increase of IoT applications, their effective scheduling in edge and cloud computing has become a critical challenge. The inherent dynamism and stochastic characteristics of edge and cloud computing, along with IoT applications, necessitate solutions that are highly adaptive. Currently, several centralized Deep Reinforcement Learning (DRL) techniques are adapted to address the scheduling problem. However, they require a large amount of experience and training time to reach a suitable solution. Moreover, many IoT applications contain multiple interdependent tasks, imposing additional constraints on the scheduling problem. To overcome these challenges, we propose a Transformer-enhanced Distributed DRL scheduling technique, called TF-DDRL, to adaptively schedule heterogeneous IoT applications. This technique follows the Actor-Critic architecture, scales efficiently to multiple distributed servers, and employs an off-policy correction method to stabilize the training process. In addition, Prioritized Experience Replay (PER) and Transformer techniques are introduced to reduce exploration costs and capture long-term dependencies for faster convergence. Extensive results of practical experiments show that TF-DDRL, compared to its counterparts, significantly reduces response time, energy consumption, monetary cost, and weighted cost by up to 60\%, 51\%, 56\%, and 58\%, respectively.
\end{abstract}


%
\IEEEpeerreviewmaketitle


\section{Introduction}
In recent years, the Internet of Things (IoT) has rapidly emerged as a transformative force, revolutionizing information technology and connectivity. The proliferation of IoT devices and applications has been exponential, reshaping the way humans interact and perceive their surroundings. Cloud computing, as a major driver of the IoT ecosystem, plays a critical role in the storage and process of the large volume of data generated by IoT devices \cite{goudarzi2022scheduling}. However, due to the potentially long physical distance between servers in cloud computing and IoT devices, high latency arises, thereby impeding the effective implementation of real-time IoT applications \cite{al2022ai}. In response to these challenges, edge computing, as a decentralized computing paradigm, has emerged to provide the ability to process, store, and intelligently control IoT applications \cite{9754225}. It has quickly become a popular computing paradigm in the IoT environment, offering substantial solutions in various domains. For instance, in the healthcare sector, edge computing can enable real-time monitoring and diagnosis, facilitating faster and more accurate medical decisions \cite{su2023cloud}; In smart cities, edge computing can be applied to real-time traffic management, improving traffic efficiency, and reducing congestion \cite{10049608}. 
\par
However, the considerable increase of IoT applications and servers within edge and cloud computing environments has brought new challenges, necessitating innovative solutions. First, there is an urgent need to minimize the expected response time of IoT applications to ensure efficient and timely performance \cite{9735306, wang2023deep}. Furthermore, in edge and cloud computing environments, the imperative need to minimize server energy consumption and monetary cost is equally crucial for sustainable and cost-effective operations \cite{pallewatta2023placement}. Thus, scheduling IoT applications on distributed servers to reduce the response time of IoT applications while simultaneously minimizing server energy consumption and monetary cost has become an important and challenging problem.
\par
Given the inherent complexity of this challenge, which can be characterized as an NP-hard problem \cite{ma2023reliability}, various solutions have been explored, including heuristic and rule-based approaches \cite{goudarzi2022scheduling}. However, these methods face limitations when dealing with the dynamic and unpredictable characteristics of servers in edge and cloud computing. The performance, utilization, and downtime of servers often lack regularity, and the number of IoT applications and their corresponding resource requirements may exhibit randomness. Additionally, IoT applications typically employ Directed Acyclic Graphs (DAGs) for modeling, where nodes represent tasks and edges signify data communication between related tasks \cite{zhou2023dag}. The dependencies between tasks introduce further complexity to the application scheduling process, rendering heuristic and rule-based solutions ineffective in addressing the scheduling challenges presented by IoT computing environments.
\par
Due to the continuous changes in edge and cloud computing environments, decision-making for scheduling IoT applications must be capable of adaptive updates. Deep Reinforcement Learning (DRL), which combines Reinforcement Learning (RL) with Deep Neural Networks (DNN), offers a promising solution. DRL agents can dynamically learn optimal policies and long-term rewards in a stochastic environment without the need for a prior understanding of the system. However, DRL agents must invest substantial time during the exploration phase by collecting extensive and diverse experience trajectories, which are later used to learn optimal policies \cite{9904958}. Hence, the effectiveness of the DRL technique can be prevented by the high exploration costs and slow convergence speeds, negatively impacting the scheduling of IoT applications in highly heterogeneous and stochastic edge and cloud computing environments.
\par
To address these challenges, we propose a distributed DRL technique, named TF-DDRL, which follows the Importance Weighted Actor-Learner Architectures (IMPALA) \cite{espeholt2018impala}, to address the problem of scheduling IoT applications in edge and cloud computing environments. TF-DDRL can effectively optimize the response time of IoT applications, server energy consumption, and total monetary cost. Also, by deploying distributed agents on various servers, TF-DDRL can expedite the generation of diverse experience trajectories, enhancing more efficient learning of policies. Moreover, the introduction of Prioritized Experience Replay (PER) \cite{schaul2015prioritized} and Transformer techniques \cite{parisotto2020stabilizing} further reduce the exploration costs and capture the long-term dependencies between features, facilitating the convergence speed of TF-DDRL. The main contributions are as follows.
\begin{itemize}
\item We propose a weighted cost model for scheduling DAG-based IoT applications in edge and cloud computing. The objective is to optimize the response time of the application, the energy consumption of the system, and the monetary cost associated with execution. Also, we customize this weighted cost model to comply with DRL algorithms.
\item We propose a distributed DRL technique, called TF-DDRL, to solve the weighted cost optimization problem. It can adaptively learn the optimal scheduling policy in response to changes in the computing environment, including diverse IoT application requests and fluctuations of computing resources. 
\item We design the network structure of TF-DDRL, integrating advanced techniques including PER and the Transformer. This design can significantly improve the convergence speed of the TF-DDRL, ensuring more efficient and effective model performance.
\item To evaluate the performance of TF-DDRL, we carry out extensive practical experiments and employ real IoT applications. Through comparisons with distributed DRL techniques including ApeX-Deep Q-Network (ApeX-DQN) \cite{horgan2018distributed} and Asynchronous Advantage Actor Critic (A3C) \cite{mnih2016asynchronous}, as well as centralized DRL techniques including Dueling Double DQN-RNN (D3QN-RNN) \cite{wang2016dueling, van2016deep} and Soft Actor-Critic (SAC) \cite{haarnoja2018soft}, we highlight the superior performance of TF-DDRL in terms of convergence speed, optimization cost, scalability, and scheduling overhead.
\end{itemize}
\par
The remainder of the paper is organized as follows. The related literature is provided in Section \ref{related_work}, and Section \ref{system_model} details the system model and formulate the scheduling problem. The main concepts of the DRL model are presented in Section \ref{drl_model}. The TF-DDRL is discussed in Section \ref{trans_ddrl}. Section \ref{performance_eva} evaluates the performance of TF-DDRL and its counterparts. Finally, the concussion and the future work are provided in Section \ref{conclusions}.
\par
\section{Related Work}
\label{related_work}
The related works that research IoT application scheduling problems in edge and cloud computing environments are studied. Related work is categorized into two groups: centralized reinforcement learning and distributed reinforcement learning.
\par
\subsection{Centralized Reinforcement Learning}
In the category of centralized reinforcement learning, Bansal et al. \cite{9933027} proposed a Dueling-DQN-based technique to place IoT applications in edge and cloud environments. It aims at optimizing the user-side latency and system energy. Hoang et al. \cite{10102429} proposed an online resource management framework based on the Actor-Critic framework, which considers the long-term constraints of queue stability and computational delay of the queuing system to minimize the average power consumption of the entire system. Huang et al. \cite{huang2019deep} focused on the resource allocation problem in edge computing environments. They developed a DQN-based approach to minimize a weighted cost, comprising total energy consumption and task completion delay. Zhao et al. \cite{10163866} proposed a mobile-aware dependent task offloading scheme based on DDPG, with the aim of minimizing the average response time and the average energy consumption of the system. Hsieh et al. \cite{10108041} investigated the task allocation problem in collaborative Mobile Edge Computing (MEC) networks, developing and comparing the performance of Double-DQN, PG, and Actor-Critic in optimizing delay and task overflow rate. The results demonstrated that the Actor-Critic approach performed the best in dynamic MEC network environments. Fan et al. \cite{10098911} studied the problem of user task offloading in MEC network environments and proposed a technique based on Dueling-DQN and Double-DQN to optimize response time and dropped task ratio. Zheng et al. \cite{9798187} defined an optimization problem involving computational offloading and resource allocation in collaborative vehicle networks. A Soft Actor-Critic (SAC)-based technique is proposed to reduce the overall delay of the system. Wang et al. \cite{8657791} proposed a computing resource allocation solution based on DQN specifically for edge computing environments. The objective of their research is to optimize the average time overhead and achieve a more balanced utilization of resources within edge environments. Xiong et al. \cite{9060882} aimed at reducing the average job completion time within edge computing environments by employing a DQN-based resource allocation policy. Jie et al. \cite{jie2021dqn} focused on the time overhead optimization problem in edge computing environments and employed a DQN-based method to reduce the task execution time. They formulate the optimization problem as a Markov Decision Process (MDP).
\par
\subsection{Distributed Reinforcement Learning}
Wen et al. \cite{wen2023fast} introduced an adaptive scheduler based on environmental changes, aiming to reduce the tail latency of edge-cloud jobs. The work employs Ape-X DQN to expedite the training process. Wang et al. \cite{9798315} proposed an Asynchronous Advantage Actor-Critic (A3C)-based approach to address the cloud-edge computing network optimization problem, aiming at satisfying the latency requirements of applications while reducing the cost of cloud servers. Garaali et al. \cite{9838831} investigated the optimization problem of computational offloading and resource allocation in an MEC environment and proposed a solution based on the A3C method. In order to reduce the system latency, each agent aims to learn the optimal offloading policy independently of the other agents in an interactive manner. Ju et al. \cite{10285560} considered the task offloading problem in vehicular edge computing networks, where the joint optimization is formulated as MDP. This work proposes an A3C-based approach to solve the MDP problem, with the goal of minimizing the system energy consumption while satisfying computational delay constraints. Sellami et al. \cite{sellami2022deep} investigated IoT application scheduling and offloading problems in edge computing environments. This work introduces a scheduling policy based on A3C to enhance energy efficiency. Chen et al. \cite{9776583} studied the task offloading problem in cloud-edge collaborative mobile computing environments, proposing an A3C-based algorithm to address the joint optimization problem involving task execution delay and energy consumption. 
Utilizing a distributed learning approach, the algorithm acquires knowledge about the probability distribution of an approximate reward and optimizes network parameters using the computing resource in the cloud, with the objective of achieving faster and more efficient decision-making.
\par
\subsection{A Qualitative Comparison}
Table \ref{tab:related_works} provides a qualitative analysis of current research work and ours in various dimensions, including application properties, architectural properties, algorithm properties, and evaluation. Application properties explore whether the IoT application has multiple tasks or not and their interdependencies. Architectural properties are divided into three layers. The IoT device layer identifies the practical characteristics of the application and the request type of the IoT devices. The section named real applications provides information on whether the work utilizes real-world IoT applications, simulations, or randomly generated data, and distinct IoT devices with varying quantities of requests and diverse requirements are classified as heterogeneous request types. The edge/cloud layer delves into the computing environment considered by the work and the heterogeneity of deployment servers. Moreover, the multi-cloud layer assesses whether the work takes into account cloud computing resources from different providers. In the algorithm properties section, the focus is on the primary techniques employed by the work and the optimization objectives. Finally, the evaluation section determines whether the work is evaluated through simulation or practical applications.
\par
In the literature, many works (e.g., \cite{10102429}, \cite{huang2019deep}, \cite{10108041}, \cite{9798187}, \cite{8657791}, \cite{9060882}, \cite{jie2021dqn}, \cite{9798315}, \cite{9838831}, \cite{10285560}, and \cite{sellami2022deep}) assume that tasks are mutually independent and do not consider the common occurrence of task dependencies in the real world. Also, works including \cite{9933027}, \cite{10102429}, \cite{huang2019deep}, \cite{10163866}, \cite{10108041}, \cite{10098911}, \cite{9798187}, \cite{8657791}, \cite{9060882}, and \cite{jie2021dqn}, adopt centralized DRL techniques, which may incur high exploration costs and exhibit low convergence speed \cite{10021988}. This poses challenges when deployed in highly distributed computing environments, especially as the number of features, environmental complexity, and application constraints increase. Furthermore, most of the work employing distributed DRL techniques is based on A3C, including \cite{9798315}, \cite{9838831}, \cite{10285560}, \cite{sellami2022deep}, and \cite{9776583}. Despite deploying distributed agents to collect experience trajectories, distributed agents in A3C train their local policies based on their limited experiences, subsequently forwarding these parameters to learners for aggregation and training, diminishing the usage efficiency of experience trajectories. To address these issues, we propose a distributed DRL technique, called TF-DDRL, which learns policies based on direct sharing of original experience, rather than parameters. Besides, TF-DDRL employs PER to enhance sampling efficiency and incorporates the Transformer to capture long-term dependencies between features, further improving convergence speed and optimizing performance. Moreover, our work considers inter-task dependencies when addressing IoT application scheduling problems in edge and multi-cloud heterogeneous environments. Also, we establish a practical experimental environment employing both real-time and non-real-time IoT applications to evaluate the performance of TF-DDRL.
\renewcommand{\arraystretch}{2}
\begin{table*}[]
\centering
\caption{A comparison of our work with existing related works}
\label{tab:related_works}
\resizebox{\textwidth}{!}{%
\begin{tabular}{|c|cc|ccccc|cccccc|c|}
\hline
\multirow{3}{*}{Work}  & \multicolumn{2}{c|}{Application Properties}                                     & \multicolumn{5}{c|}{Architectural Properties}                                                                                                                                                 & \multicolumn{6}{c|}{Algorithm Properties}                                                                                                                                                                                          & \multirow{3}{*}{Evaluation} \\ \cline{2-14}
                       & \multicolumn{1}{c|}{\multirow{2}{*}{Task Number}} & \multirow{2}{*}{Dependency} & \multicolumn{2}{c|}{IoT Device Layer}                                  & \multicolumn{2}{c|}{Edge/Cloud Layer}                                           & \multirow{2}{*}{Multi-Cloud Layer} & \multicolumn{2}{c|}{\multirow{2}{*}{Main Technique}}                                                       & \multicolumn{4}{c|}{Optimization Objectives}                                                                          &                             \\ \cline{4-7} \cline{11-14}
                       & \multicolumn{1}{c|}{}                             &                             & \multicolumn{1}{c|}{Real Applications} & \multicolumn{1}{c|}{Request Type}  & \multicolumn{1}{c|}{Computing Environemnt} & \multicolumn{1}{c|}{Heterogeneity} &                                    & \multicolumn{2}{c|}{}                                                                                      & \multicolumn{1}{c|}{Time}       & \multicolumn{1}{c|}{Energy}     & \multicolumn{1}{c|}{Finance}    & Multi Objective &                             \\ \hline
\cite{9933027}         & \multicolumn{1}{c|}{Multiple}                     & Dependent                   & \multicolumn{1}{c|}{\LEFTcircle}  & \multicolumn{1}{c|}{Heterogeneous} & \multicolumn{1}{c|}{Edge and Cloud}        & \multicolumn{1}{c|}{Heterogeneous} & $\times$                           & \multicolumn{1}{c|}{\multirow{10}{*}{Centralized}} & \multicolumn{1}{c|}{Dueling-DQN}                      & \multicolumn{1}{c|}{\checkmark} & \multicolumn{1}{c|}{\checkmark} & \multicolumn{1}{c|}{$\times$}   & \checkmark      & Simulation                  \\ \cline{1-8} \cline{10-15} 
\cite{10102429}        & \multicolumn{1}{c|}{Single}                       & Independent                 & \multicolumn{1}{c|}{\Circle}      & \multicolumn{1}{c|}{Homogeneous}   & \multicolumn{1}{c|}{Edge}                  & \multicolumn{1}{c|}{Heterogeneous} & $\times$                           & \multicolumn{1}{c|}{}                              & \multicolumn{1}{c|}{Actor-Critic}                     & \multicolumn{1}{c|}{$\times$}   & \multicolumn{1}{c|}{\checkmark} & \multicolumn{1}{c|}{$\times$}   & $\times$        & Simulation                  \\ \cline{1-8} \cline{10-15} 
\cite{huang2019deep}    & \multicolumn{1}{c|}{Multiple}                     & Independent                 & \multicolumn{1}{c|}{\LEFTcircle}  & \multicolumn{1}{c|}{Heterogeneous} & \multicolumn{1}{c|}{Edge}                  & \multicolumn{1}{c|}{Homogeneous}   & $\times$                           & \multicolumn{1}{c|}{}                              & \multicolumn{1}{c|}{DQN}                              & \multicolumn{1}{c|}{\checkmark} & \multicolumn{1}{c|}{\checkmark} & \multicolumn{1}{c|}{$\times$}   & \checkmark      & Simulation                  \\ \cline{1-8} \cline{10-15} 
\cite{10163866}        & \multicolumn{1}{c|}{Multiple}                     & Dependent                   & \multicolumn{1}{c|}{\LEFTcircle}  & \multicolumn{1}{c|}{Heterogeneous} & \multicolumn{1}{c|}{Edge}                  & \multicolumn{1}{c|}{Heterogeneous} & $\times$                           & \multicolumn{1}{c|}{}                              & \multicolumn{1}{c|}{DDPG}                             & \multicolumn{1}{c|}{\checkmark} & \multicolumn{1}{c|}{\checkmark} & \multicolumn{1}{c|}{$\times$}   & \checkmark      & Simulation                  \\ \cline{1-8} \cline{10-15} 
\cite{10108041}        & \multicolumn{1}{c|}{Single}                       & Independent                 & \multicolumn{1}{c|}{\LEFTcircle}  & \multicolumn{1}{c|}{Heterogeneous} & \multicolumn{1}{c|}{Edge and Cloud}        & \multicolumn{1}{c|}{Heterogeneous} & $\times$                           & \multicolumn{1}{c|}{}                              & \multicolumn{1}{c|}{\begin{tabular}[c]{@{}c@{}}Double-DQN, PG, \\ and Actor-Critic\end{tabular}} & \multicolumn{1}{c|}{\checkmark} & \multicolumn{1}{c|}{$\times$}   & \multicolumn{1}{c|}{$\times$}   & $\times$        & Simulation                  \\ \cline{1-8} \cline{10-15} 
\cite{10098911}        & \multicolumn{1}{c|}{Multiple}                     & Dependent                   & \multicolumn{1}{c|}{\Circle}      & \multicolumn{1}{c|}{Homogeneous}   & \multicolumn{1}{c|}{Edge}                  & \multicolumn{1}{c|}{Homogeneous}   & $\times$                           & \multicolumn{1}{c|}{}                              & \multicolumn{1}{c|}{\begin{tabular}[c]{@{}c@{}}Dueling-DQN \\ and Double-DQN\end{tabular}}       & \multicolumn{1}{c|}{\checkmark} & \multicolumn{1}{c|}{$\times$}   & \multicolumn{1}{c|}{$\times$}   & \checkmark      & Simulation                  \\ \cline{1-8} \cline{10-15} 
\cite{9798187}         & \multicolumn{1}{c|}{Single}                       & Independent                 & \multicolumn{1}{c|}{\LEFTcircle}  & \multicolumn{1}{c|}{Homogeneous}   & \multicolumn{1}{c|}{Edge}                  & \multicolumn{1}{c|}{Homogeneous}   & $\times$                           & \multicolumn{1}{c|}{}                              & \multicolumn{1}{c|}{SAC}                              & \multicolumn{1}{c|}{\checkmark} & \multicolumn{1}{c|}{$\times$}   & \multicolumn{1}{c|}{$\times$}   & $\times$        & Simulation                  \\ \cline{1-8} \cline{10-15} 
\cite{8657791}         & \multicolumn{1}{c|}{Single}                       & Independent                 & \multicolumn{1}{c|}{\LEFTcircle}  & \multicolumn{1}{c|}{Homogeneous}   & \multicolumn{1}{c|}{Edge}                  & \multicolumn{1}{c|}{Homogeneous}   & $\times$                           & \multicolumn{1}{c|}{}                              & \multicolumn{1}{c|}{DQN}                              & \multicolumn{1}{c|}{\checkmark} & \multicolumn{1}{c|}{$\times$}   & \multicolumn{1}{c|}{$\times$}   & \checkmark      & Simulation                  \\ \cline{1-8} \cline{10-15} 
\cite{9060882}         & \multicolumn{1}{c|}{Multiple}                     & Independent                 & \multicolumn{1}{c|}{\LEFTcircle}  & \multicolumn{1}{c|}{Homogeneous}   & \multicolumn{1}{c|}{Edge}                  & \multicolumn{1}{c|}{Homogeneous}   & $\times$                           & \multicolumn{1}{c|}{}                              & \multicolumn{1}{c|}{DQN}                              & \multicolumn{1}{c|}{\checkmark} & \multicolumn{1}{c|}{$\times$}   & \multicolumn{1}{c|}{$\times$}   & $\times$        & Simulation                  \\ \cline{1-8} \cline{10-15} 
\cite{jie2021dqn}      & \multicolumn{1}{c|}{Single}                       & Independent                 & \multicolumn{1}{c|}{\LEFTcircle}  & \multicolumn{1}{c|}{Homogeneous}   & \multicolumn{1}{c|}{Edge}                  & \multicolumn{1}{c|}{Homogeneous}   & $\times$                           & \multicolumn{1}{c|}{}                              & \multicolumn{1}{c|}{DQN}                              & \multicolumn{1}{c|}{\checkmark} & \multicolumn{1}{c|}{$\times$}   & \multicolumn{1}{c|}{$\times$}   & $\times$        & Simulation                  \\ \hline
\cite{wen2023fast}     & \multicolumn{1}{c|}{Multiple}                     & Dependent                   & \multicolumn{1}{c|}{\LEFTcircle}  & \multicolumn{1}{c|}{Heterogeneous} & \multicolumn{1}{c|}{Edge and Cloud}        & \multicolumn{1}{c|}{Heterogeneous} & $\times$                           & \multicolumn{1}{c|}{\multirow{7}{*}{Distributed}}  & \multicolumn{1}{c|}{Ape-X DQN}                        & \multicolumn{1}{c|}{\checkmark} & \multicolumn{1}{c|}{$\times$}   & \multicolumn{1}{c|}{$\times$}   & $\times$        & Simulation                  \\ \cline{1-8} \cline{10-15} 
\cite{9798315}         & \multicolumn{1}{c|}{Single}                       & Independent                 & \multicolumn{1}{c|}{\Circle}      & \multicolumn{1}{c|}{Homogeneous}   & \multicolumn{1}{c|}{Edge and Cloud}        & \multicolumn{1}{c|}{Homogeneous}   & $\times$                           & \multicolumn{1}{c|}{}                              & \multicolumn{1}{c|}{A3C}                              & \multicolumn{1}{c|}{\checkmark} & \multicolumn{1}{c|}{$\times$}   & \multicolumn{1}{c|}{\checkmark} & \checkmark      & Simulation                  \\ \cline{1-8} \cline{10-15} 
\cite{9838831}         & \multicolumn{1}{c|}{Multiple}                     & Independent                 & \multicolumn{1}{c|}{\Circle}      & \multicolumn{1}{c|}{Homogeneous}   & \multicolumn{1}{c|}{Edge}                  & \multicolumn{1}{c|}{Homogeneous}   & $\times$                           & \multicolumn{1}{c|}{}                              & \multicolumn{1}{c|}{A3C}                              & \multicolumn{1}{c|}{\checkmark} & \multicolumn{1}{c|}{$\times$}   & \multicolumn{1}{c|}{$\times$}   & $\times$        & Simulation                  \\ \cline{1-8} \cline{10-15} 
\cite{10285560}        & \multicolumn{1}{c|}{Single}                       & Independent                 & \multicolumn{1}{c|}{\LEFTcircle}  & \multicolumn{1}{c|}{Heterogeneous} & \multicolumn{1}{c|}{Edge}                  & \multicolumn{1}{c|}{Heterogeneous} & $\times$                           & \multicolumn{1}{c|}{}                              & \multicolumn{1}{c|}{A3C}                              & \multicolumn{1}{c|}{$\times$}   & \multicolumn{1}{c|}{\checkmark} & \multicolumn{1}{c|}{$\times$}   & $\times$        & Simulation                  \\ \cline{1-8} \cline{10-15} 
\cite{sellami2022deep} & \multicolumn{1}{c|}{Single}                       & Independent                 & \multicolumn{1}{c|}{\LEFTcircle}  & \multicolumn{1}{c|}{Heterogeneous} & \multicolumn{1}{c|}{Edge}                  & \multicolumn{1}{c|}{Heterogeneous} & $\times$                           & \multicolumn{1}{c|}{}                              & \multicolumn{1}{c|}{A3C}                              & \multicolumn{1}{c|}{$\times$}   & \multicolumn{1}{c|}{\checkmark} & \multicolumn{1}{c|}{$\times$}   & $\times$        & Simulation                  \\ \cline{1-8} \cline{10-15} 
\cite{9776583}         & \multicolumn{1}{c|}{Multiple}                     & Dependent                   & \multicolumn{1}{c|}{\Circle}      & \multicolumn{1}{c|}{Homogeneous}   & \multicolumn{1}{c|}{Edge and Cloud}        & \multicolumn{1}{c|}{Heterogeneous} & $\times$                           & \multicolumn{1}{c|}{}                              & \multicolumn{1}{c|}{A3C}                              & \multicolumn{1}{c|}{\checkmark} & \multicolumn{1}{c|}{\checkmark} & \multicolumn{1}{c|}{$\times$}   & \checkmark      & Simulation                  \\ \cline{1-8} \cline{10-15} 
Our work               & \multicolumn{1}{c|}{Multiple}                     & Dependent                   & \multicolumn{1}{c|}{\CIRCLE}      & \multicolumn{1}{c|}{Heterogeneous} & \multicolumn{1}{c|}{Edge and Cloud}        & \multicolumn{1}{c|}{Heterogeneous} & \checkmark                         & \multicolumn{1}{c|}{}                              & \multicolumn{1}{c|}{IMPALA}                           & \multicolumn{1}{c|}{\checkmark} & \multicolumn{1}{c|}{\checkmark} & \multicolumn{1}{c|}{\checkmark} & \checkmark      & Practical                   \\ \hline
\multicolumn{14}{l}{\CIRCLE: Real IoT Application and Deployment, \LEFTcircle: Simulated IoT Application, \Circle: Random} 
\end{tabular}%
}
\end{table*}
\renewcommand{\arraystretch}{1}
\par
\section{System Model and Problem Formulation}
\label{system_model}
This section first describes the topology of IoT systems in this work. Next, we tackle the scheduling of IoT applications by formulating it as an optimization problem, aiming at reducing application response time, system energy consumption, and monetary cost of running applications. Table \ref{table:notations} depicts the key notations used in this paper.
\begin{table*}[]
\caption{List of key notations}
\label{table:notations}
\resizebox{\textwidth}{!}{%
\begin{tabular}{ll|ll}
\hline
\textbf{Variable}                                        & \textbf{Description}                                                                                    & \textbf{Variable}         & \textbf{Description}                                             \\ \hline
$\mathcal{A}$                                            & One application set                                                                                     & $PR(\mathcal{A}^j_i)$     & The set of predecessor tasks of task $\mathcal{A}^j_i$           \\
$\mathcal{A}_i$                                          & One application                                                                          & $SU(\mathcal{A}^j_{i})$   & The set of successor tasks of task $\mathcal{A}^j_i$             \\
$\mathcal{A}_i^j$                                        & One single task                                                                                         & $T$                       & The response time model                                          \\
$N$                                                      & The server set                                                                                          & $T^{dat}$                 & The Data Arrival Time (DAT) model                                \\
$\mathcal{CS}$                                           & The cloud server set                                                                                    & $T^{ex}$                  & The execution time model                                         \\
$\mathcal{ES}$                                           & The edge server set                                                                                     & $T^{tr}$                  & The transmission time model                                      \\
$N_k$                                                    & One single server                                                                                       & $E$                       & The energy consumption model                                     \\
$Freq(\mathcal{N}_k)$                                    & The CPU frequency (MHz) of server $\mathcal{N}_k$                                                       & $E^{ex}$                  & The execution energy model                                       \\
$Ram(\mathcal{N}_k)$                                     & The RAM size (GB) of server $\mathcal{N}_k$                                                             & $E^{tr}$                  & The transmission energy model                                    \\
$\mathcal{C}_i$                                          & The scheduling configuration for application $\mathcal{A}_i$                                            & $W^{ex}(\mathcal{N}_{k})$ & The power of the server $\mathcal{N}_{k}$ when executing task    \\
$\mathcal{C}_i^j$                                        & The scheduling configuration for task $\mathcal{A}_i^j$                                                 & $W^{tr}(\mathcal{N}_{k})$ & The power of the server $\mathcal{N}_{k}$ when transmitting data \\
$\mathcal{P}_{\mathcal{N}_l, \mathcal{N}_k}$             & The propagation time (ms) between server $\mathcal{N}_l$ and server $\mathcal{N}_k$                     & $F$                       & The monetary cost model                                          \\
$DS_{\mathcal{N}_{l}, \mathcal{N}_{k}}(\mathcal{A}^j_i)$ & The data size for task $\mathcal{A}^j_i$ sent from server $\mathcal{N}_{l}$ to server $\mathcal{N}_{k}$ & $CLP(\mathcal{N}_{k})$    & The pricing of the cloud server $\mathcal{N}_{k}$                \\
$\mathcal{B}_{\mathcal{N}_l, \mathcal{N}_k}$             & The data rate (bits/second) between server $\mathcal{N}_l$ and server $\mathcal{N}_k$                   & $EP(\mathcal{N}_{k})$     & The electricity price for running edge server $\mathcal{N}_{k}$  \\
$L(\mathcal{A}^j_i)$                                     & The required CPU cycles for task $\mathcal{A}^j_i$                                                      & $J$                       & The weighted cost model                 \\ \hline                        
\end{tabular}%
}
\end{table*}
\par
\subsection{System Model}
\begin{figure}[!htb]
\centering
\includegraphics[width=\linewidth]{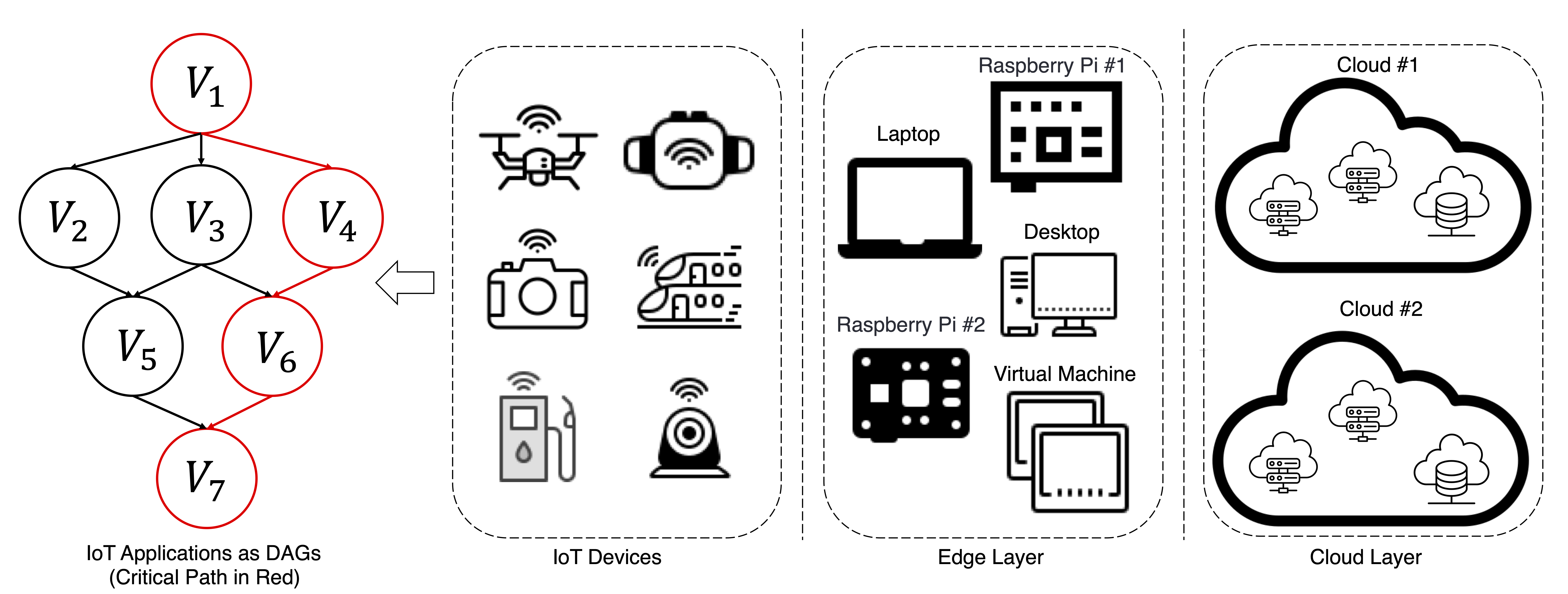}
\caption{An overview of Edge and Cloud computing}
\label{fig:ovw}
\end{figure}

Fig.~\ref{fig:ovw} provides a layered perspective of the IoT system in an edge and cloud environment. Consider $\mathcal{A} = \{\mathcal{A}_i|1 \leq i \leq |\mathcal{A}|\}$ as a collection of $|\mathcal{A}|$ applications, with each application comprising one or more tasks denoted as $\mathcal{A}_i = \{\mathcal{A}^j_i|1 \leq j \leq |\mathcal{A}_i|\}$. To model an IoT application, we use a DAG, as illustrated in Fig.~\ref{fig:ovw}, where each vertex $\mathcal{V}_j = \mathcal{A}^j_i$ corresponds to a specific task within the application $\mathcal{A}_i$. The edges, represented as $\mathcal{E}_{j,k}$, signify the data flow between tasks $\mathcal{V}_j$ and $\mathcal{V}_k$, indicating that successor tasks must follow the completion of their predecessors. Also, the critical path of the DAG, denoted as $CP(\mathcal{A}_i)$ and marked in red in the figure, shows the path with the highest cost.
\par
We consider a server set comprising $|\mathcal{N}|$ servers to handle the application set $\mathcal{A}$, denoted as $\mathcal{N} = \mathcal{CS} \cup \mathcal{ES}\ = \{\mathcal{N}_k|1 \leq k \leq |\mathcal{N}|\}$. $\mathcal{CS}$ denotes the cloud server set and $\mathcal{ES}$ denotes the edge server set. To consider server heterogeneity, each server $\mathcal{N}_k$ is characterized by different CPU frequency (MHz) $Freq(\mathcal{N}_k)$ and RAM size (GB) $Ram(\mathcal{N}_k)$. Also, we consider $\mathcal{P}_{\mathcal{N}_l, \mathcal{N}_k}$ as the propagation time (ms) and $\mathcal{B}_{\mathcal{N}_l, \mathcal{N}_k}$ as the data rate (b/s) between server $\mathcal{N}_l$ and $\mathcal{N}_k$.
\par
\subsection{Problem Formulation}
\label{problem formulation}
Since an application consists of one or more tasks, it can run on various servers. Considering server set as $\mathcal{N}$, the scheduling configuration $\mathcal{C}^j_{i}$ for task $\mathcal{A}^j_i$ is defined as:
\begin{equation}
\mathcal{C}^j_{i} = \mathcal{N}_k\,
\end{equation}
where $k$ shows the server index. The scheduling configuration $\mathcal{C}_{i}$ for the application $\mathcal{A}_i$ is a collection of the scheduling configurations for the tasks within $\mathcal{A}_i$, and is defined as:
\begin{equation}
\mathcal{C}_{i} = \{\mathcal{C}^j_{i} | 1 \leq j \leq |\mathcal{A}_i|\}.
\end{equation}
\par
Also, the task execution model of one application can exhibit hybrid characteristics, incorporating both sequential and/or parallel processes. Accordingly, each task cannot be executed unless all predecessor tasks complete their execution, while tasks that are not dependent on each other can be executed in parallel. We use $PR(\mathcal{A}^j_i)$ to denote the set of predecessor tasks of task $\mathcal{A}^j_i$ and use $CP(\mathcal{A}^j_i)$ to indicate whether task $\mathcal{A}^j_i$ is located on the critical path of the application $\mathcal{A}_i$.
\par
\subsubsection{Response Time Model}
Assuming that the scheduling configuration for task $\mathcal{A}^j_i$ is $\mathcal{C}^j_{i}$, the response time model $T(\mathcal{C}^j_{i})$ consists of two parts, the Data Arrival Time (DAT) model $T^{dat}(\mathcal{C}^j_{i})$ and the execution time model $T^{ex}(\mathcal{C}^j_{i})$:
\begin{equation}
T(\mathcal{C}^j_{i}) =  T^{dat}(\mathcal{C}^j_{i}) + T^{ex}(\mathcal{C}^j_{i}). 
\end{equation}
The DAT model $T^{dat}(\mathcal{C}^j_{i})$ signifies the maximum time for the data, required by task $\mathcal{A}^j_i$, to reach the designated server:
\begin{equation}
T^{dat}(\mathcal{C}^j_{i}) = max\;\;T_{\mathcal{C}^k_{i}, \mathcal{C}^j_{i}}^{dat}, \quad \forall \mathcal{A}^k_{i}\in PR(\mathcal{A}^j_i),
\end{equation}
where $T_{\mathcal{C}^k_{i}, \mathcal{C}^j_{i}}^{dat}$ shows the time consumed for the required data to be transmitted from scheduled server $\mathcal{C}^k_{i}$ to server $\mathcal{C}^j_{i}$. Here, $\mathcal{C}^j_{i}$ signifies the server scheduled for the execution of task $\mathcal{A}^j_i$, while $\mathcal{C}^k_{i}$ corresponds to the server where the predecessor task of task $\mathcal{A}^j_i$ is executed. Thus, $T_{\mathcal{C}^k_{i}, \mathcal{C}^j_{i}}^{dat}$ depends on both the transmission time $T_{\mathcal{C}^k_{i}, \mathcal{C}^j_{i}}^{tr}$ and the propagation time $\mathcal{P}_{\mathcal{C}^k_{i}, \mathcal{C}^j_{i}}$ for task $\mathcal{A}^j_i$ between server $\mathcal{C}^k_{i}$ and server $\mathcal{C}^j_{i}$:
\begin{equation} 
T_{\mathcal{C}^k_{i}, \mathcal{C}^j_{i}}^{dat} = 
\begin{cases}
T_{\mathcal{C}^k_{i}, \mathcal{C}^j_{i}}^{tr} + \mathcal{P}_{\mathcal{C}^k_{i}, \mathcal{C}^j_{i}}& \mathcal{C}^k_{i} \neq \mathcal{C}^j_{i},\\
0& \mathcal{C}^k_{i} = \mathcal{C}^j_{i}.
\end{cases}
\end{equation}
where the transmission time $T_{\mathcal{C}^k_{i}, \mathcal{C}^j_{i}}^{tr}$ is calculated as follows:
\begin{equation}
T_{\mathcal{C}^k_{i}, \mathcal{C}^j_{i}}^{tr} = \frac{DS_{\mathcal{C}^k_{i}, \mathcal{C}^j_{i}}(\mathcal{A}^j_i)}{\mathcal{B}_{\mathcal{C}^k_{i}, \mathcal{C}^j_{i}}},
\end{equation}
$DS_{\mathcal{C}^k_{i}, \mathcal{C}^j_{i}}(\mathcal{A}^j_i)$ denotes the data size for task $\mathcal{A}^j_i$ sent from server $\mathcal{C}^k_{i}$ to server $\mathcal{C}^j_{i}$, and $\mathcal{B}_{\mathcal{C}^k_{i}, \mathcal{C}^j_{i}}$ denotes the current bandwidth between server $\mathcal{C}^k_{i}$ and server $\mathcal{C}^j_{i}$.
\par
The execution time model $T^{ex}(\mathcal{C}^j_{i})$ is defined as the time required to execute task $\mathcal{A}^j_i$ based on scheduling configuration $\mathcal{C}^j_{i}$. It can be calculated as follows:
\begin{equation}\label{eq:etm}
T^{ex}(\mathcal{C}^j_{i}) = \frac{L(\mathcal{A}^j_i)}{Freq(\mathcal{C}^j_{i})}, 
\end{equation}
where $L(\mathcal{A}^j_i)$ denotes the necessary CPU cycles for task $\mathcal{A}^j_i$ to be executed and $Freq(\mathcal{C}^j_{i})$ shows the CPU frequency of scheduled server $\mathcal{C}^j_{i}$ (if the CPU has multiple cores, the model considers the average frequency). Accordingly, the formulation of the response time model $T(\mathcal{C}_{i})$ for application $\mathcal{A}_i$ is expressed as follows:
\begin{equation}\label{eq:tts}
T(\mathcal{C}_{i}) = \sum_{j=1}^{|\mathcal{A}_i|}(T(\mathcal{C}^j_{i}) \times CP(\mathcal{A}^j_i)), 
\end{equation}
where $CP(\mathcal{A}^j_i)$ is the critical path indicator. If task $\mathcal{A}^j_i$ belongs to the critical path of application $\mathcal{A}_i$, $CP(\mathcal{A}^j_i)$ is set to $1$; otherwise, it assumes a value of $0$.
\par
\subsubsection{Energy Consumption Model}
In this work, we consider the energy consumption of the edge layer and the cloud layer. Given that the scheduling configuration for task $\mathcal{A}^j_i$ is $\mathcal{C}^j_{i}$, the energy consumption $E(\mathcal{C}^j_{i})$ is determined by the energy consumed during the actual task processing (i.e., the execution energy model) $E^{ex}(\mathcal{C}^j_{i})$, plus the energy consumed by the servers when transmitting the required data to other servers (i.e., transmission energy model) $E^{tr}(\mathcal{C}^j_{i})$:
\begin{equation}
E(\mathcal{C}^j_{i}) = E^{ex}(\mathcal{C}^j_{i}) + (E^{tr}(\mathcal{C}^j_{i}) \times ED(\mathcal{A}^j_i)),
\end{equation}
where $ED(\mathcal{A}^j_i)$ is $0$ if $\mathcal{A}^j_i$ is the ending task (i.e., has no successor task) in application $\mathcal{A}_i$ and $1$ otherwise.
\par
The execution energy model $E^{ex}(\mathcal{C}^j_{i})$ is the energy consumed by the server to execute the task, defined as:
\begin{equation}
E^{ex}(\mathcal{C}^j_{i}) = T^{ex}(\mathcal{C}^j_{i}) \times W^{ex}(\mathcal{C}^j_{i}), 
\end{equation}
where $T^{ex}(\mathcal{C}^j_{i})$ is obtained from Eq. \ref{eq:etm} and $W^{ex}(\mathcal{C}^j_{i})$ represents the power of the server when executing the task.
\par
Considering the dependency between tasks, one task $\mathcal{A}^j_i$ can have one or more predecessor tasks. The transmission energy model $E^{tr}(\mathcal{C}^j_{i})$ is defined as the sum of the energy consumed to transmit the data to the servers where the successor tasks are assigned, as follows: 
\begin{equation}\label{eq:tre}
\begin{array}{l}
E^{tr}(\mathcal{C}^j_{i}) = 
\sum\limits_{\mathcal{A}^l_{i} \in SU(\mathcal{A}^j_{i})}\frac{DS_{\mathcal{C}^j_{i}, \mathcal{C}^l_{i}}(\mathcal{A}^j_i)}{\mathcal{B}_{\mathcal{C}^j_{i}, \mathcal{C}^l_{i}}} \times W^{tr}(\mathcal{C}^j_{i}) \\
\times OS(\mathcal{C}^j_{i}, \mathcal{C}^l_{i}), 
\end{array}
\end{equation}
where $SU(\mathcal{A}^j_{i})$ denotes the set of successor tasks of task $\mathcal{A}^j_i$, $\mathcal{C}^l_{i}$ is the scheduling configuration of task $\mathcal{A}^l_{i}$, $W^{tr}(\mathcal{C}^j_{i})$ show the transmission power of the server when transmitting data, and $OS(\mathcal{C}^j_{i}, \mathcal{C}^l_{i})$ is 0 if $\mathcal{C}^j_{i}$ and $\mathcal{C}^j_{i}$ are the same server and 1 otherwise. Similar to \cite{7248815}, \cite{long2018energy}, $W^{tr}(\mathcal{C}^j_{i})$ is set as a constant value, but this parameter can also be dynamically adjusted.
\par
Accordingly, the energy consumption model $E(\mathcal{C}_{i})$ for application $\mathcal{C}_{i}$ is formulated as follows:
\begin{equation}\label{eq:ets}
E(\mathcal{C}_{i}) = \sum_{j=1}^{|\mathcal{A}_i|}E(\mathcal{C}^j_{i}), 
\end{equation}
\par
\subsubsection{Monetary Cost Model}
The server set $\mathcal{N}$ comprises both the cloud server set $\mathcal{CS}$ and the edge server set $\mathcal{ES}$. Without loss of generality, we assume that the edge servers are on-premises servers and are owned by users, so their execution cost only depends on their electricity usage. Otherwise, the cloud-like pricing model can be used for edge servers.  Given that the scheduling configuration for the task $\mathcal{A}^j_i$ is $\mathcal{C}^j_{i}$, the monetary cost $F(\mathcal{C}^j_{i})$ depends both on the cloud server and the edge server price models. Formally, the monetary cost model $F(\mathcal{C}^j_{i})$ is:
\begin{equation}
F(\mathcal{C}^j_{i}) = 
\begin{cases}
\sum\limits_{\mathcal{C}^j_{i} \in \mathcal{CS}}T(\mathcal{C}^j_{i}) \times CLP(\mathcal{C}^j_{i})  & \mathcal{C}^j_{i} \in \mathcal{CS},\\ \\
\sum\limits_{\mathcal{C}^j_{i} \in \mathcal{ES}}E(\mathcal{C}^j_{i}) \times EP(\mathcal{C}^j_{i}) & \mathcal{C}^j_{i} \in \mathcal{ES},
\end{cases}
\end{equation}
where $CLP(\mathcal{C}^j_{i})$ shows the cloud server pricing, and $EP(\mathcal{C}^j_{i})$ denotes the electricity price for running edge server $\mathcal{C}^j_{i}$. Consequently, the monetary cost model $F(\mathcal{C}_{i})$ for the application $\mathcal{C}_{i}$ is formulated as follows:
\begin{equation}\label{eq:fts}
F(\mathcal{C}_{i}) = \sum_{j=1}^{|\mathcal{A}_i|}F(\mathcal{C}^j_{i}).\end{equation}
\par
\subsubsection{Weighted Cost Model}
\label{wcm}
The weighted cost model $J(\mathcal{C}^j_i)$ is defined as the weighted sum of the normalized response time models, the energy consumption model, and the monetary cost model. Given the scheduling configuration for task $\mathcal{A}^j_i$ is $\mathcal{C}^j_{i}$:
\begin{equation}\label{eq:wt}
\begin{array}{l}
J(\mathcal{C}^j_i) = w_1\frac{T(\mathcal{C}^j_i) - T^{min}}{T^{max} - T^{min}} + w_2\frac{E(\mathcal{C}^j_i) - E^{min}}{E^{max} - E^{min}} + \\ w_3\frac{F(\mathcal{C}^j_i) - F^{min}}{F^{max} - F^{min}},
\end{array}
\end{equation}
where $T^{min}$, $T^{max}$, $E^{min}$, $E^{max}$, $F^{min}$, and $F^{max}$ represent the minimum and the maximum value that can be achieved by the response time model, the energy consumption model, and the monetary cost model, respectively. Also, $w_1$, $w_2$, and $w_3$ are the control parameters used to fine-tune the weighted cost model. The reason for employing normalized models, rather than the original models, is that the values of the models may fall within different ranges.
\par
Accordingly, the weighted cost model for application $\mathcal{A}_i$ is defined as:
\begin{equation}\label{eq:final}
\hspace{-0.5cm}
\begin{array}{l}
J(\mathcal{C}_i) = w_1 \times Norm(T(\mathcal{C}_i)) + w_2 \times Norm(E(\mathcal{C}_i)) + \\
w_3 \times Norm(F(\mathcal{C}_i)),
\end{array}
\end{equation}
where $T(\mathcal{C}_i)$, $E(\mathcal{C}_i)$, and $F(\mathcal{C}_i)$ are obtained from Eq. \ref{eq:tts} , Eq. \ref{eq:ets} and Eq. \ref{eq:fts}, and $Norm$ represents the normalization. 
\par
Therefore, the optimization problem of scheduling IoT applications can be formulated as:
\begin{align}
\hspace{-5cm}
\label{eq:was}
     min \quad & J(\mathcal{C}_i) \\ 
  \text{s.t.} \quad 
     & C1:\;Size(\mathcal{C}^j_i) = 1,\;\forall \mathcal{C}^j_i \in \mathcal{C}_i \\
     & C2:\;DS_{\mathcal{C}^k_{i}, \mathcal{C}^j_{i}}(\mathcal{A}^j_i), \mathcal{B}_{\mathcal{C}^k_{i}, \mathcal{C}^j_{i}} > 0,\; \forall \mathcal{C}^k_{i}, \mathcal{C}^j_{i} \in \mathcal{N}, \notag\\
     & \qquad\;\;\forall \mathcal{A}^j_{i} \in \mathcal{A}_{i} \\
     & C3:\;Freq(\mathcal{N}_k), Ram(\mathcal{N}_k) > 0,\; \forall \mathcal{N}_k \in \mathcal{N}\\
     & C4:\;\sum_{\mathcal{A}_i \in \mathcal{A}}\sum_{\mathcal{A}^j_i \in \mathcal{A}_i}Ram(\mathcal{A}^j_i) \times SO(\mathcal{A}^j_i, \mathcal{N}_k)<Ram(\mathcal{N}_k), \notag \\
     & \qquad\;\;\forall \mathcal{N}_k\in \mathcal{N} \\
     & C5:\;T(\mathcal{A}^j_i) \leq T(\mathcal{A}^j_i + \mathcal{A}^k_i),\forall \mathcal{A}^j_i \in PR(\mathcal{A}^k_i) \\
     & C6:\;w_1+w_2+w_3=1,\; 0 \leq w_1,w_2,w_3 \leq 1
\end{align}
where $C1$ enforces the rule that each task can be assigned to only one server. $C2$ specifies the transmission constraints for data size and bandwidth. Additionally, $C3$ defines constraints related to the CPU frequency and RAM size of the server. Furthermore, $C4$ ensures that every server has adequate RAM resources to process all tasks scheduled on it. $SO(\mathcal{A}^j_i, \mathcal{N}_k)$ equals 1 if task $\mathcal{A}^j_i$ is scheduled on server $\mathcal{N}_k$, otherwise 0. $C5$ specifies that each task is eligible for processing only after the completion of its predecessor tasks, ensuring that the accumulative cost is no less than that of the predecessor task. Lastly, $C6$ places restrictions on the control parameters within Eq. \ref{eq:final}, confining them to values between $0$ and $1$.
\par
The problem under consideration is characterized as a non-convex optimization problem, primarily due to the potential existence of an infinite number of local optima within the feasible domain. Typically, algorithms aimed at finding the global optimum in such problems exhibit exponential complexity and are classified as NP-hard \cite{qiu2020distributed}. To solve such a non-convex optimization problem, most approaches decompose these problems into several convex sub-problems \cite{ding2021budget}, subsequently solving these sub-problems iteratively until convergence is achieved \cite{tran2019federated}. However, this strategy often sacrifices accuracy for reduced complexity \cite{ji2022trajectory}. Also, these approaches are heavily dependent on the current environment and are not suitable for dynamic environments with highly heterogeneous computational resources \cite{goudarzi2023mu}. To tackle this issue, we propose TF-DDRL to adaptively manage uncertainties in dynamic and stochastic environments. It can dynamically learn scheduling policies through continuous interaction with the environment.
\par
\subsection{Deep Reinforcement Learning Model}
\label{drl_model}
To apply the DRL approach, the optimization problem should be formulated as a MDP. More specifically, the problem can be defined by the tuple $<\mathbb{S},\mathbb{A},\mathbb{P},\mathbb{R},\gamma>$, where
$\mathbb{S}$ signifies a finite set of states, $\mathbb{A}$ represents a finite set of actions, $\mathbb{P}$ represents the state transition probability, $\mathbb{R}$ stands for the reward function, and $\gamma \in [0,1]$ serves as the discount factor employed in calculating cumulative rewards.
\par
We consider the learning process to be divided into multiple time steps $t$ within a total time span $\mathbb{T}$. At each time step, the agent interacts with the environment, resulting in multiple states $S_t$. At time step $t$, the agent observes the environment state $S_t = s$, where $s \in \mathbb{S}$. Guided by the policy $\pi(a|s)$, where $a \in \mathbb{A}$, the agent chooses an action $A_t = a$. The policy function $\pi(a|s) = Pr[A_t=a|S_t=s]$ explicitly defines the probability of selecting action $a$ given state $s$. Following the execution of action $a$, the agent receives a reward $r=\mathbb{R}[S_t = s, A_t = a]$ from the environment, determined by the reward function $\mathbb{R}$. The agent then undergoes a state transition to $S_{t+1} = s'$ based on the state transition function $P^a_{ss'} = \mathbb{P}[S_{t+1} = s'|S_t = s, A_t = a]$. The ultimate objective of the agent is to acquire a policy $\pi$ maximizing the expected cumulative discounted reward, denoted as $\mathbb{E}\pi[\sum{t \in T}\gamma_tr_t]$.
\par
Considering the scheduling problem of IoT applications in edge and cloud computing environments, the MDP's state space $\mathbb{S}$, action space $\mathbb{A}$, and reward function $\mathbb{R}$ are defined as follows:
\begin{itemize}
\item \textbf{State space $\mathbb{S}$}: 
In this work, the formulated problem pertains to tasks and servers, with the state $\mathbb{S}$ containing $\mathbb{F}$ for the task feature and $\mathbb{G}$ for the server set state. At time step $t$, the feature space $\mathbb{F}$ of task $\mathcal{A}^j_i$ includes essential details including task ID, predecessors, successors, application ID, etc, defined as:
\begin{equation}
 \mathbb{F}_t(\mathcal{A}^j_i) = \{f_t^y(\mathcal{A}^j_i)| \mathcal{A}^j_i \in \mathcal{A}_i, 0 \leq y \leq |\mathbb{F}|\},
\end{equation}
where $y$ denotes the feature index and $|\mathbb{F|}$ represents the total number of features. Also, in time step $t$, the state space $\mathbb{G}$ of the server set $\mathcal{N}$ contains the number of servers, the CPU frequency, RAM size, label (e.g., cloud or edge), expense per time unit (for cloud servers), electricity price (for edge servers), propagation time, bandwidth between different servers, etc, which is formally defined as:
\begin{equation}
\begin{array}{l}
\mathbb{G}_t(\mathcal{N}) = \{|\mathcal{N}|, g_t^z(\mathcal{N}_k), h_t^q(\mathcal{N}_j, \mathcal{N}_k)| \\
\mathcal{N}_j, \mathcal{N}_k \in \mathcal{N}, 0 \leq z \leq |g|, 0 \leq q \leq |h|\},
\end{array}
\end{equation}
where $g$ is the sub-state set containing states associated with an individual server (e.g., CPU utilization), and $z$ corresponds to its index. Additionally, $h$ signifies the sub-state set containing states associated with two servers (e.g., propagation time), and $q$ denotes the index. Consequently, $\mathbb{S}$ is defined as:
\begin{equation}
 \mathbb{S} = \{S_t = (\mathbb{F}_t(\mathcal{A}^j_i), \mathbb{G}_t(\mathcal{N}))|\mathcal{A}^j_i \in \mathcal{A}_i, t \in \mathbb{T}\}.
\end{equation}
\item \textbf{Action space $\mathbb{A}$}: 
In this work, scheduling involves the action of assigning the current task $\mathcal{A}^j_i$ to an individual server $\mathcal{N}_k$. Consequently, the definition of the action at time step $t$ is as follows:
\begin{equation}
A_t = \mathcal{C}^j_i = \mathcal{N}_k.
\end{equation}
Therefore, the action space $\mathbb{A}$ equals to the server set $N$:
\begin{equation}
\mathbb{A} = \mathcal{N}.
\end{equation}
\item \textbf{Reward function $\mathbb{R}$}: 
As outlined in Section \ref{wcm}, the primary objective is to minimize the weighted cost model presented in Eq. \ref{eq:was}. Thus, in time step $t$, the reward $r_t$ can be defined as the negative value of Eq. \ref{eq:wt} if the task can be successfully executed. However, if the task $\mathcal{A}^j_i$ fails to be executed on the scheduled server $\mathcal{C}^j_i$, a substantial negative value is introduced as a penalty. Formally, $r_t$ is defined as:
\begin{equation}
\label{eq:rf}
r_t =
\begin{cases}
- J(\mathcal{C}^j_i) & succeed \\
penalty& fail,
\end{cases}
\end{equation}
\end{itemize}
\par
\section{TF-DDRL: Distributed DRL Framework}
\label{trans_ddrl}
\begin{figure}[!htb]
\centering
\includegraphics[width=0.5\linewidth]{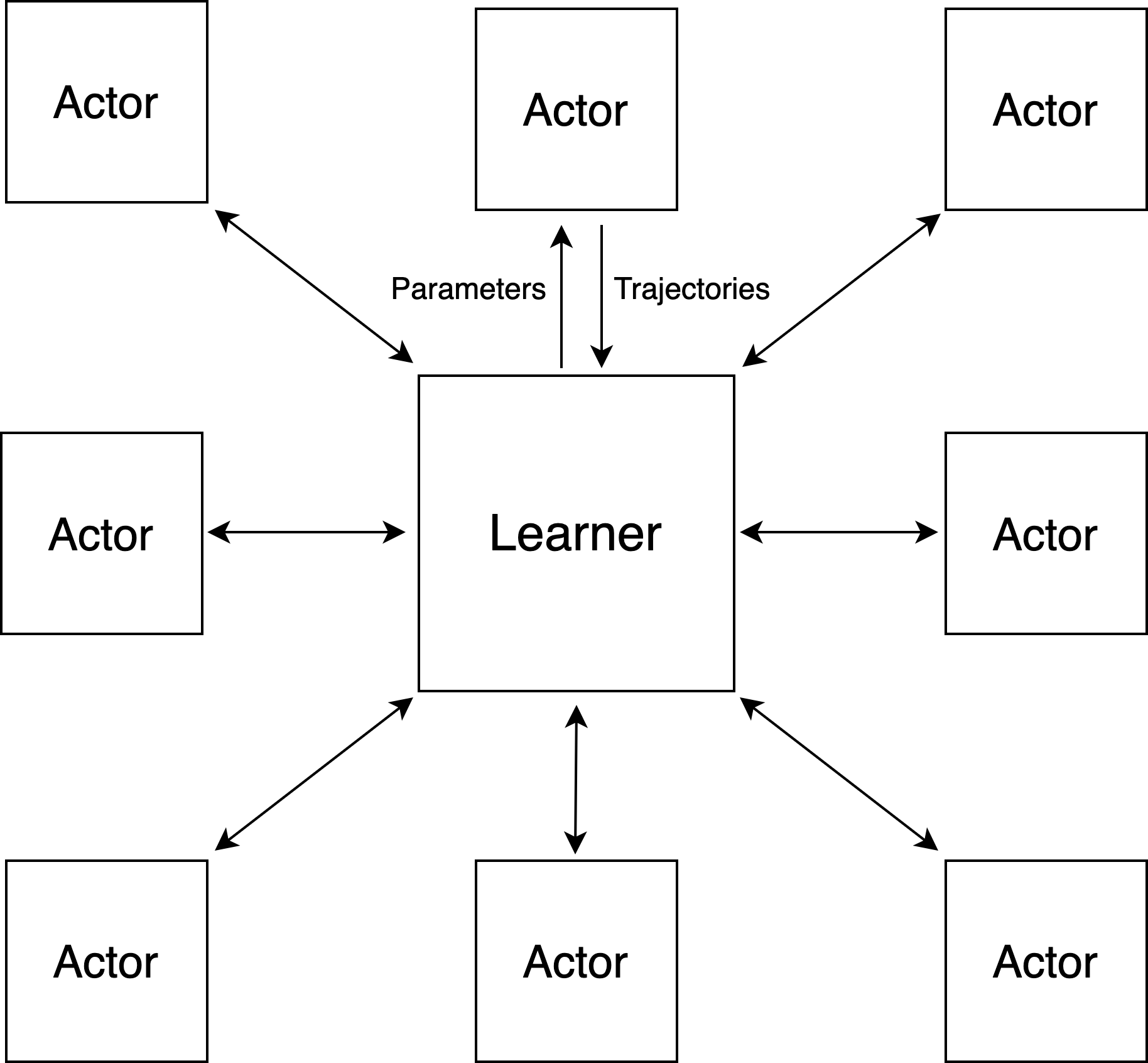}
\caption{High-level architecture of Actors and Learner}
\label{fig:hla}
\end{figure}
\begin{figure}[!htb]
\centering
\includegraphics[width=0.75\linewidth]{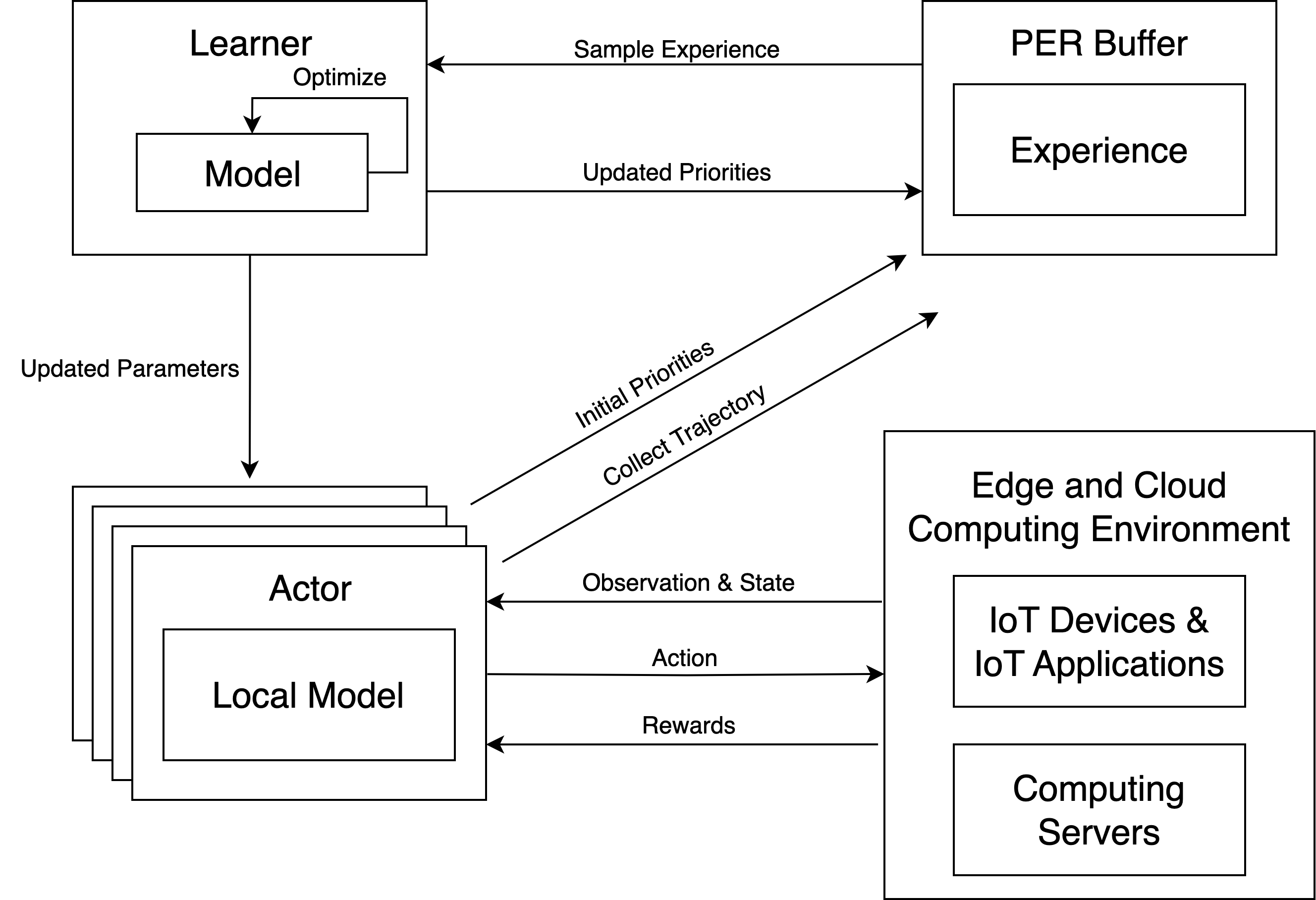}
\caption{An overview of TF-DDRL framework}
\label{fig:tdf}
\end{figure}

The high-level architecture of the TF-DDRL framework is depicted in Fig. \ref{fig:hla}. The architecture comprises multiple Actors responsible for collecting data to create experience trajectories and a Learner that leverages the experience trajectories to learn a policy $\pi$. The primary objective is to identify a policy $\pi$ that maximizes the expected sum of future discounted rewards: 
\begin{equation}\label{eq:ob}
V^{\pi}(s) = \mathbb{E}_{\pi}[\sum_{t \in T}\gamma_tr_t], 
\end{equation}
where $\pi$ represents the policy, $\gamma \in [0, 1]$ is the discount factor, $r_t = r(s_t, a_t)$ denotes the reward at time $t$, $s_t$ is the state at time $t$, $s$ is the initial state $s_0$, and $a_t = \pi(a_t|s_t)$ is the action generated by following a specific policy $\pi$. Fig. \ref{fig:tdf} presents an overview of the TF-DDRL framework. In what follows, each component and communication process is discussed in detail.
\par
\subsection{Actor: Experience Trajectories Generation}
Algorithm \ref{ag:actor} describes how the Actor in the TF-DDRL framework generates experience trajectories. In order to improve the efficiency of sampling and the speed of convergence of TF-DDRL, PER is introduced to store the trajectory experiences of the Actor. At the beginning, the Actor updates its local policy $\mu$ to the most recent Learner policy $\pi$ and initializes a PER buffer $\mathcal{P}$ to store the collected transitions. Before one trajectory, the Actor generates the initial state based on the information of the current task and server set. After that, based on the output $a_t$ of the policy $\mu$, the Actor schedules the current task to the corresponding server. Then, the reward $r_t$ of the current action $a_t$ is calculated based on Eq. \ref{eq:rf}, and the next state $s_{t+1}$ is also generated based on the information of the next task and the server set. Afterward, the Actor calculates the importance measure $m_t$ and stores the current transition $(s_t, a_t, r_t, \mu(a_t|s_t),s_{t+1})$ into $\mathcal{P}$ based on $m_t$. After $n$ steps, the Actor sends the trajectory $\{s_1, a_1, r_1, \mu(a_1|s_1),...,s_n, a_n, r_n, \mu(a_n|s_n), s_{n+1}\}$ to the Learner. The Learner then iteratively updates its policy $\pi$ over a batch of trajectories gathered from different Actors. This framework decouples data collection and learning, allowing more Actors to be added and distributed across multiple machines for efficient utilization of computing resources in edge and cloud IoT systems. 
\begin{algorithm}
\footnotesize
\caption{Actor: experience generation}\label{ag:actor}
\SetKwInOut{Input}{Input}
\SetKwInOut{Initialization}{Initialization}
\Input{the Actors's local policy $\mu$; the Learner's policy $\pi$; the Learner's address $Learner$; max time step $n$;} 
\While{True}{
    $\mu \gets UpdateActorPolicy(\pi, Learner)$\;
    $\mathcal{P} \gets InitializePERBuffer()$\;
    $servers \gets GetServers()$\;
    $task \gets GetTask()$\;
    $s_1 \gets GenerateState(servers, task)$\;
    \For{$t \gets x$ to $x+n-1$}{
        $a_t \gets \mu(s_t)$\;
        $Schedule(task, a_t)$\;
        $r_t \gets GetReward()$\; 
        $servers \gets GetServers()$\;
        $task \gets GetTask()$\;
        $s_{t+1} \gets GenerateState(servers, task)$\;
        $e_t = (s_t, a_t, r_t, \mu(a_t|s_t),s_{t+1})$\;
        $m_t = r_t + \gamma_t V(s_{t+1}) - V(s_t)$\;
        Store transition $e_t$ into $\mathcal{P}$ based on $m_t$\;
    }
    \If{$Length(\mathcal{P}) == n$}{
        $SubmitTrajectory(\mathcal{P}, Learner)$\;
    }
}
\end{algorithm}
\par
\subsection{Learner: Schedule Policy Update}
However, it's worth noting that after a few updates, the Actor's strategy $\mu$ may fall behind the Learner's strategy $\pi$. To address the gap between the Actor's policy $\mu$ and the Learner's policy $\pi$, an off-policy correction method named V-trace \cite{espeholt2018impala} is introduced to rectify this discrepancy. 
\par
\subsubsection{V-trace Correction Method}
The Learner in TF-DDRL maintains a state value function $V$ based on the samples from the Actors. The purpose of the V-trace correction method is to provide an estimate of the current state value function $V$, called V-trace target $\Hat{V}$. After $n$ steps of interaction with the environment, the Actor collects a trajectory $(s_t, a_t, r_t, \mu(a_t|s_t), s_{t+1})^{t=x+n}_{t=x}$ following its policy $\mu$. The n-steps V-trace target $\Hat{V}(s_x)$ for state $s_x$ is: 
\begin{equation}\label{eq:vtrace}
\Hat{V}(s_x) = V(s_x) + \sum^{x+n-1}_{t=x}\gamma^{t-x}(\prod^{t-1}_{i=x}c_i)\delta_tV,
\end{equation}
where $\delta_tV$ is the truncated Temporal Difference (TD) for $V$, and $\prod^{t-1}_{i=x}c_i$ measures the impact of $\delta_tV$ observed at time $t$ on the update of the value function $V$ at the previous time $x$. Specifically, $\delta_tV$ is defined as:
\begin{equation}
\delta_tV = \rho_t(r_t+\gamma_tV(s_{t+1}))-V(s_t)),
\end{equation}
and $c_i$ and $\rho_t$ are truncated importance sampling weights,:
\begin{equation}
c_i = min (\bar{c}, \frac{\pi(a_i|s_i)}{\mu(a_i|s_i)}),
\end{equation}
\begin{equation}\label{eq:pt}
\rho_t = min (\bar{\rho}, \frac{\pi(a_t|s_t)}{\mu(a_t|s_t)}),
\end{equation}
where $\bar{c}$ and $\bar{\rho}$ are the truncation constants with $\bar{c} \leq \bar{\rho}$. $\bar{c}$ affects the speed of convergence, while $\bar{\rho}$ affects the solution to which the value function $V$ converges. Considering $\bar{\rho}$, the corresponding target policy $\pi_{\bar{\rho}}(a|x)$ is defines as:
\begin{equation}
\pi_{\bar{\rho}}(a|x) = \frac{min(\bar{\rho}\mu(a|x), \pi(a|x))}{\sum_{b \in A}min(\bar{\rho}\mu(b|x), \pi(b|x))}
\end{equation}
\par
\subsubsection{Actor-Critic-based Algorithm}
The implementation of TF-DDRL follows the Actor-Critic architecture. TF-DDRL optimizes two DNNs, the actor (policy) network and the critic (value) network. The actor network focuses on acquiring a policy $\pi$ to maximize the expected cumulative discounted reward $\mathbb{E}_\pi[\sum_{t \in T}\gamma_tr_t]$. Meanwhile, the critic network evaluates the current policy $\pi$ by computing the TD error, which measures the difference between the current reward and the estimate of the value function $V$.
\par
Algorithm \ref{ag:ddrl} describes how the Learner in the TF-DDRL framework updates policies. The Learner first obtains the collected trajectories from all Actors. In order to improve the efficiency of sampling and the speed of convergence of the algorithm, trajectory experiences are sampled based on importance measure $m_x$. When updating the networks, the loss function of TF-DDRL is defined as follows: 
\begin{equation}
\begin{array}{l}
loss_{total} = a_v*loss_{value}+a_p*loss_{policy}+ \\
a_e*loss_{entropy},
\end{array}
\end{equation}
where $loss_{value}$ is the loss function for value function, $loss_{policy}$ is the loss function for policy, $loss_{entropy}$ is the loss function for entropy bonus, and $a_v, a_p$, and $a_e$ are the corresponding weights. Considering $\pi_{\phi}$ is the current policy parameterized by $\phi$, $V_{\theta}$ is the value function parameterized by $\theta$, and $\mu$ is the Actor's local policy, the value loss function $loss_{value}$ is defined as the $L_2$ loss between the current value $V_{\theta}$ and the V-trace target value $\Hat{V}$:
\begin{equation}
loss_{value} = (\Hat{V}(s_x)-V_{\theta}(s_x))^2,
\end{equation}
where $\Hat{V}(s_x)$ is from Eq. \ref{eq:vtrace}. Considering the objective function Eq. \ref{eq:ob}, the policy gradient can be presented as:
\begin{equation}
\nabla V^{\pi}(s) = \mathbb{E}_{\pi}[\nabla log \pi(a_x|s_x) Q_\pi(s_x,a_x)],
\end{equation}
where $Q_\pi(s_x,a_x)$ is the state-value of policy $\pi$ at $(s_x,a_x)$. In TF-DDRL, the truncated importance sampling weight $\rho_x$ between the policy $\pi_{\bar{\rho}}$ and the Actor's local policy $\mu$ is employed to suppress the divergence. Also, we use $r_s + \gamma v_{s+1}$, named as the v-trace advantage, to estimate $Q_{\pi_{\bar{\rho}}}(a_x|s_x)$. Besides, state-dependent baseline $V_{\theta}(s_x)$ is subtracted from the v-trace advantage to reduce bias. Therefore, the policy loss function $loss_{policy}$ is defined as:
\begin{equation}
loss_{policy} = -\rho_x log\pi_{\phi}(a_x|s_x)(r_x+\gamma v_{x+1}-V_{\theta}(s_x)),
\end{equation}
where $\rho_x$ is from Eq. \ref{eq:pt}. We also exploit the entropy $H(\pi_{\phi})$ as a bonus to encourage exploration, with the loss function $loss_{entropy}$ defined as:
\begin{equation}
loss_{entropy} = -H(\pi_{\phi}) = \sum_{a}\pi_{\phi}(a_x|s_x)log\pi_{\phi}(a_x|s_x)
\end{equation}
Therefore, the value function parameter $\theta$ is updated in the direction of:
\begin{equation}
\Delta \theta = a_v*(\Hat{V}(s_x)-V_{\theta}(s_x))\nabla_{\theta}V_{\theta}(s_x),
\end{equation}
and the policy parameter $\phi$ is updated through policy gradient: 
\begin{equation}
\begin{split}
\Delta \phi = a_p\rho_x\nabla_{\phi}log\pi_{\phi}(a_x|s_x)(r_x+\gamma v_{s+1}-V_{\theta}(s_x)) \\
- a_e\nabla_{\phi}\sum_{a}\pi_{\phi}(a_x|s_x)log\pi_{\phi}(a_x|s_x).
\end{split}
\end{equation}
\par
\begin{algorithm}
\footnotesize
\caption{Learner: policy update}\label{ag:ddrl}
\SetKwInOut{Input}{Input}
\SetKwInOut{Initialization}{Initialization}
\Input{current policy $\pi_{\phi}$; value function $V_{\theta}$; update epoch $X$; buffer size $N$; value function loss coefficient $a_v$; policy objective function loss coefficient $a_c$; entropy bonus loss coefficient $a_e$; the Actors set $Actors$} 
\While{True}{
    $\mathcal{D} \gets InitializeBuffer()$\;
    \For{$actor$ in $Actors$}{
        $\mathcal{D}.append(ReceiveTrajectory(actor))$\;
    }
    \For{$trajectory$ in $\mathcal{D}$}{
        \For{$x \gets 1$ to $X$}{
            Sample experience $e_x = (s_x, a_x, r_x, \mu(a_x|s_x),s_{x+1}) \sim \mathcal{M}(x) = m_{x}^{\alpha}/\sum_{i}m^{\alpha}_{i}$\;
            $w_x = (N\mathcal{M}(x))^{-\beta}/max_iw_i$\;
            $\Hat{V}(s_x) \gets V_{\theta}(s_x) + \sum^{x+n-1}_{t=x}\gamma^{t-x}(\prod^{t-1}_{i=x}c_i)\delta_tV_{\theta}$\;
            $m_x \gets |\delta_t|$\;
            $loss_{value} \gets (\Hat{V}(s_x)-V_{\theta}(s_x))^2$\;
            $loss_{policy} \gets -\rho_xlog\pi_{\phi}(a_x|s_x)(r_x+\gamma v_{x+1}-V_{\theta}(s_x))$\;
            $loss_{entropy} \gets \sum_{a}\pi_{\phi}(a_x|s_x)log\pi_{\phi}(a_x|s_x)$\;
            $loss_{total} \gets a_v*loss_{value}+a_p*loss_{policy}+a_e*loss_{entropy}$\;
            $\Delta \theta \gets \Delta \theta + w_x a_v(\Hat{V}(s_x)-V_{\theta}(s_x))\nabla_{\theta}V_{\theta}(s_x)$\;
            $\Delta \phi \gets \Delta \phi + w_x (a_p\rho_x\nabla_{\phi}log\pi_{\phi}(a_x|s_x)(r_x+\gamma v_{s+1}-V_{\theta}(s_x)) - a_e\nabla_{\phi}\sum_{a}\pi_{\phi}(a_x|s_x)log\pi_{\phi}(a_x|s_x))$\;
        }
        update $\theta$ and $\phi$ by Adam optimizer\;
    }
    $BroadcastPolicy(Actors, \pi_{\phi})$\;
}
\end{algorithm}
\par
\subsection{Prioritized Experience Replay}
The Learner in TF-DDRL relies on the experience trajectories collected from Actors to update the parameters. However, in dynamic edge and cloud environments, the experience changes over time, resulting in significant gaps between samples. Each sample can contribute to different improvements to the model. To enhance sampling efficiency, expedite convergence, and enable the model to quickly adapt to changes by focusing on the most pertinent experiences during the training phase, PER \cite{schaul2015prioritized} is introduced in both the data collection phase and the model update phase.
\par
As presented in Algorithm \ref{ag:actor}, during the data collection phase, the Actor assigns importance measure $m_t$ to each experience sample when storing it in the buffer. Since the TD error reflects the difference between the model's estimated value of the current state and the next state, and when this difference is significant, it indicates that the experience sample provides valuable information for updating the current policy. Therefore, TF-DDRL uses the TD error as a metric to measure the importance of samples, defined as:
\begin{equation}
m_t = r_t + \gamma_t V(s_{t+1}) - V(s_t) + \epsilon, 
\end{equation}
where $\epsilon$ is a tiny positive number from 0 to 1, in case the experience is not sampled when the TD error is 0.
As presented in Algorithm \ref{ag:ddrl}, when the Learner samples experience $e_x$ from the trajectory, the sampling probability $\mathcal{M}(x)$ is calculated as follows:
\begin{equation}
e_x \sim \mathcal{M}(x) = \frac{m_{x}^{\alpha}}{\sum_{i}m^{\alpha}_{i}}
\end{equation}
where $\alpha$ determines the degree of priority, and $\alpha = 0$ corresponds to the uniform case (i.e., each experience has the same probability of being sampled). 
\par
However, when experiences are given priority, they have different probabilities of being sampled, which will introduce bias in the update of the value network, thus changing the direction of the convergence of the value network. In order to correct this error, an importance-sampling weight is added to each empirical sample, calculated as follows:
\begin{equation}
w_x = (\frac{1}{N\mathcal{M}(x)})^{\beta}*(max_iw_i)^{-1},
\end{equation}
where $N$ is the number of experience samples in the PER buffer, $\beta$ is a hyperparameter within $0$ and $1$ that will gradually increase and finally settle at 1, and $(max_iw_i)^{-1}$ is to normalize the weight to improve stability. The purpose of using the importance-sampling weight is to strike a balance between prioritizing samples to learn important experiences and reducing the potential bias. As $\beta$ continues to rise to 1, the bias gradually decreases, and the learning process gradually reduces the impact of prioritization, ensuring a more stable and unbiased learning process. This helps prevent the model from becoming too sensitive to specific experiences, encouraging a more robust and accurate learning process.
\par
\subsection{Gated Transformer-XL}
\begin{figure}[!htb]
\centering
\includegraphics[width=0.75\linewidth, height=5cm]{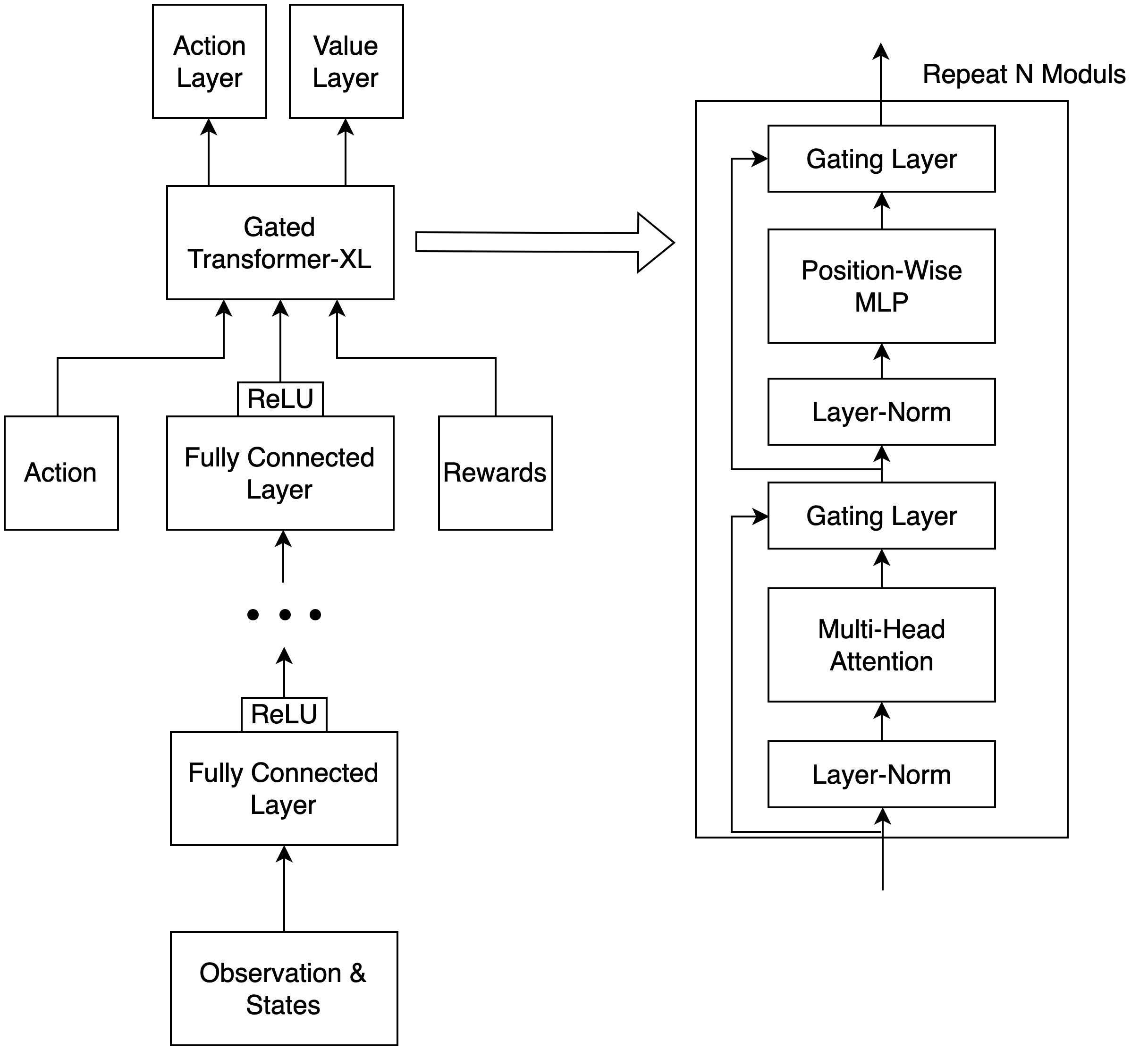}
\caption{The network architecture of TF-DDRL framework}
\label{fig:nta}
\end{figure}

Due to the heterogeneity and dynamics of edge and cloud environments, TF-DDRL uses the Gated Transformer-XL \cite{parisotto2020stabilizing} to allow the model to better capture long-term dependencies and global relationships between states. The network architecture of TF-DDRL is shown in Fig. \ref{fig:nta}.
\par
In the Transformer layer of TF-DDRL, the Multi-Head Attention block applies the attention mechanism to different linear mappings (heads) of the input and concatenates them together, allowing the model to focus on different parts of the input sequence simultaneously, which helps to capture the relationships between different features in the input. The Position-wise Multi-Layer Perceptron (MLP) block is used to perform independent nonlinear transformations of features at each position, enhancing the model's ability to capture complex patterns and relationships at different positions in the input sequence, providing a more expressive representation for downstream processing. The Gating Layer is used to weight the features of each position when passing through different blocks. The model can control the flow of input information by learning the appropriate weights, making it more suitable for specific tasks and data distribution.
\par
\section{Performance Evaluation}
\label{performance_eva}
This section introduces the experiment configuration, hyperparameters of the TF-DDRL, and the performance study.
\par
\subsection{Experiment Setup}
We discuss the specification of our practical edge-cloud environment, details of employed real IoT applications, and baseline techniques.
\par
\subsubsection{Practical Experiment Environment}
To reflect the heterogeneous computing environments, a practical experiment environment, containing IoT devices, edge servers, and cloud servers, is established. Besides, to build a multi-cloud computing environment, we used three instances from the Nectar Cloud infrastructure (All AMD EPYC with 2 cores @2.0GHz, 8GB RAM; 4cores @2.0GHz, 16GB RAM; 8 cores @2.0GHz, 32GB RAM), one instance from AWS Cloud (Intel Xeon with 1 core @2.4GHz, 1GB RAM), and one instance from Microsoft Azure Cloud (Intel Xeon with 1 core @2.3GHz, 1GB RAM).
\par
In the edge computing environment, we used one RPi 3B (Pi OS, Broadcom BCM2837 with 4 cores @1.2GHz, 1GB RAM), one Macbook Pro (macOS, M1 Pro with 8 cores, 16GB RAM), and one Dell laptop (Linux, Intel Core i7 with 8 cores @2.3GHz, 16GB RAM). Also, as IoT devices, we have used webcams, IP cameras, and docker containers that stream pre-recorded video files.
\par
Moreover, we used the Victorian Default Offer\footnote{https://www.esc.vic.gov.au/electricity-and-gas/prices-tariffs-and-benchmarks/victorian-default-offer} (i.e., 0.2871 AUD/kWh) in Australia as the electricity price, and the official price of AWS\footnote{https://aws.amazon.com/pricing} and Microsoft Azure\footnote{https://azure.microsoft.com/pricing} cloud servers (i.e., 0.1296 AUD/hour for m6a.large, 0.2592 AUD/hour for m6a.xlarge, 0.5184 AUD/hour for m6a.2xlarge, 0.0174 AUD/hour for t2.micro, and 0.0156 AUD/hour for B1s) to calculate the monetary cost in the experiment. In our environment, the servers exhibit the following average latency and bandwidth (data rate): the latency between the IoT device and the Nectar cloud server ranges from 6-12ms, with a bandwidth between 14-20MB/s; between the IoT device and the AWS cloud server, the latency ranges from 15-25ms, with a bandwidth between 15-22MB/s; between the IoT device and the Microsoft Azure cloud server, the latency ranges from 7-15ms, with a bandwidth between 15-21MB/s; the latency between the IoT device and the edge servers ranges from 1-6ms, with a bandwidth between 130-140 MB/s. The energy consumed to execute applications on servers is monitored using the eco2AI library \cite{budennyy2022eco2ai}. Moreover, the transmission power $W^{tr}(\mathcal{C}^j_{i})$ of servers is obtained similar to \cite{7248815}, \cite{long2018energy}, and  $W^{tr}(\mathcal{C}^j_{i})$ is set between 0.75-1 watt for edge servers and between 3-5 watt for cloud servers. However, these parameters can be adjusted.
\par
Furthermore, in Eq. \ref{eq:final}, $w_1$, $w_2$, and $w_3$ are set to 0.33, indicating that the importance of response time, energy consumption, and monetary cost are considered equal.
\par
\subsubsection{Sample IoT Applications}
\label{app}
To evaluate TF-DDRL's performance, we utilized four IoT applications, featuring real-time and/or non-real-time capabilities. Real-time functionality allows applications to process live streams, while non-real-time functionality facilitates the processing of pre-recorded video files. The applications adhere to a sensor-actuator architecture. Also, all applications offer an adjustable parameter known as the \textit{application label}, which determines the resolution of the video. The applications are detailed below:
\begin{itemize}
\item \textit{Face Detection} \cite{goudarzi2021resource}: Identifies human faces in real-time, marking them with squares in the video. This application is implemented using the OpenCV\footnote{https://github.com/opencv/opencv\label{opencv}}.
\item \textit{Color Tracking} \cite{goudarzi2021resource}:  Traces colors in a video stream in real-time. Users have the flexibility to dynamically configure target colors using the application's GUI. This application is developed using OpenCV\footref{opencv}.
\item \textit{Face And Eye Detection} \cite{goudarzi2021resource}: Alongside identifying human faces in real-time, it detects human eyes. This application is developed using OpenCV\footref{opencv}.
\item \textit{Video OCR} \cite{deng2021fogbus2}: Retrieves textual content from pre-recorded video and presents it to the user. It is designed to automatically filter keyframes for efficient processing. This application is developed using Google’s Tesseract-OCR Engine\footnote{https://github.com/tesseract-ocr/tesseract}.
\end{itemize}
\par
\subsubsection{Baseline Techniques}
To evaluate the performance of TF-DDRL, we implemented five additional DRL techniques, including centralized and distributed, as outlined below:
\begin{itemize}
\item \textit{IMPALA} \cite{espeholt2018impala}: It is a distributed DRL technique and is designed for large-scale environments. TF-DDRL is based on the architecture of IMPALA to enable high scalability in highly distributed environments.
\item \textit{ApeX-DQN} \cite{horgan2018distributed}: It is an improved DRL technique based on DQN that introduces a distributed learning architecture, adopted by Wen et al. \cite{wen2023fast} for scheduling problems.
\item \textit{A3C} \cite{mnih2016asynchronous}: It is one of the most adapted techniques in the distributed DRL field for scheduling problems. It has been used by many works in the current literature, including \cite{9798315}, \cite{9838831}, \cite{10285560}, \cite{sellami2022deep}, and \cite{9776583}. It combines the Actor-Critic method with the concept of concurrent execution. We extend this technique to solve the proposed optimization problem in the heterogeneous edge and cloud computing environment.
\item \textit{D3QN-RNN}: Many works  (\cite{9933027}, \cite{huang2019deep}, \cite{10108041}, \cite{10098911}, \cite{9060882}, and \cite{jie2021dqn}) use DQN-based DRL techniques. We extend the foundation of DQN, incorporating the Dueling architecture \cite{wang2016dueling} to decompose the Q values into state and advantage values for a more accurate estimation of the relative value of actions. Also, we introduce Double DQN \cite{van2016deep}, employing two independent neural networks to estimate target Q-values to address the overestimation during training. Moreover, RNN is used in this technique.
\item \textit{SAC} \cite{haarnoja2018soft}: It is a centralized DRL technique and is used by Zheng et al. \cite{9798187}. It combines the Actor-Critic method with entropy regularization, encouraging exploration, and enhancing the stability of learning. It is extended to address our problem within the heterogeneous edge and cloud computing environment.
\end{itemize}
\par
\subsection{Technique Hyperparameters}
The network architecture of TF-DDRL is depicted in Fig. \ref{fig:nta}. In our implementation, we used three fully connected layers, followed by two Gated Transformer-XL-based attention layers, and then two additional fully connected layers for generating action logits and the value function. Furthermore, we performed a grid search to fine-tune the hyperparameters. Accordingly, we set the learning rate ($lr$) to 0.001 and the discount factor ($\gamma$) to 0.99. Also, the $\Bar{c}$ and $\Bar{\rho}$, governing the V-trace performance, are both set to 1 for optimal results. Table \ref{table:hyperparameters} provides a summary of the hyperparameter settings. Moreover, we conducted hyperparameter tuning for the baseline techniques to ensure a fair assessment of their performance, as presented in Table \ref{table:hyperparameters_baseline}.
\begin{table}[]
\footnotesize
\caption{The hyperparameters setting for TF-DDRL}
\label{table:hyperparameters}
\centering
\begin{tabular}{|l|l|}
\hline
\textbf{TF-DDRL Hyperparameter}                       & \textbf{Value} \\ \hline
Fully Connected Layers                            & 3              \\ \hline
Hidden Layer Units                       & [256, 256, 128]             \\ \hline
Activation Function                      & ReLU           \\ \hline
Learning Rate $lr$                     & 0.001         \\ \hline
Discount Factor $\gamma$                          & 0.99            \\ \hline
Transformer Unit Number                 & 2        \\ \hline
Transformer Head Number                 & 4        \\ \hline
Transformer Head Dimension              & 32        \\ \hline
Transformer Position-wise MLP Dimension              & 32        \\ \hline
Optimization Method                      & Adam           \\ \hline
\end{tabular}%
\end{table}
\par
\begin{table}[]
\centering
\caption{Hyperparameters of baseline techniques}
\label{table:hyperparameters_baseline}
\resizebox{\linewidth}{!}{%
\begin{tabular}{|l|l|l|l|l|}
\hline
\textbf{Hyperparameters} & \textbf{ApeX-DQN} & \textbf{A3C}      & \textbf{D3QN-RNN} & \textbf{SAC}               \\ \hline
Fully Connected Layers   & 3                 & 3                 & 3                 & 3                 \\ \hline
Hidden Layer Units       & {[}256,256,128{]} & {[}256,256,128{]} & {[}256,256,128{]} & {[}256,256,128{]} \\ \hline
Activation Function      & ReLU              & TanH              & ReLU              & ReLU              \\ \hline
Learning Rate            & 0.001             & 0.001             & 0.001             & 0.0001            \\ \hline
Discount Factor          & 0.99              & 0.9               & 0.99              & 0.99              \\ \hline
\end{tabular}%
}
\end{table}
\par
\subsection{Performance Study}
The results of our extensive experiments are shown below.
\par
\subsubsection{PER and Transformer Analysis}
This experiment studies the performance of TF-DDRL compared to native IMPALA. We employ the four applications detailed in Section \ref{app} for training. Due to the page limit, the results are provided exclusively for weighted costs.
\par
\begin{figure}[!htb]
\centering
\includegraphics[width=0.45\textwidth, height=3.5cm]{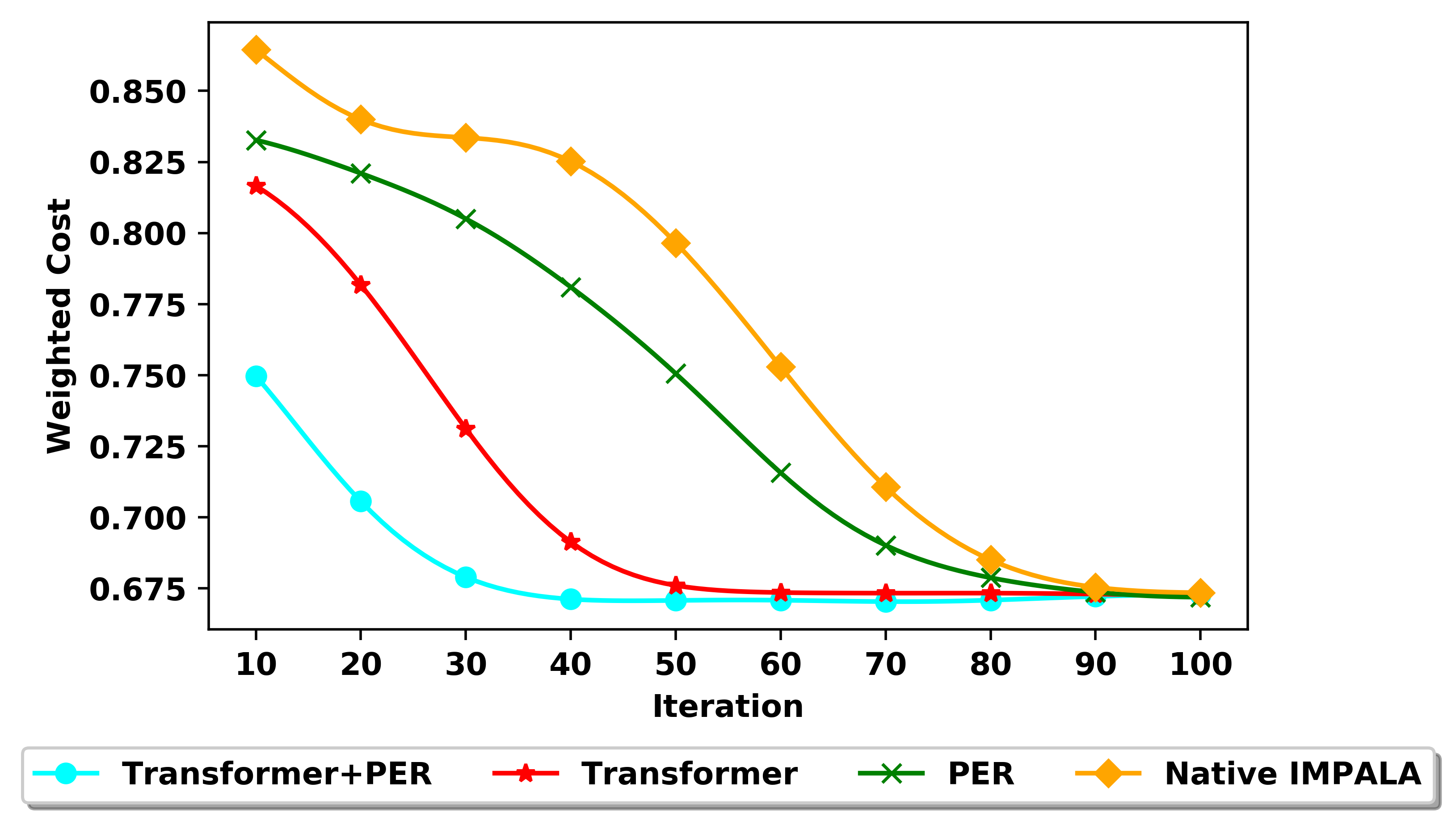}
\caption{PER and Transformer analysis}
\label{fig:fla}
\end{figure}

Figure \ref{fig:fla} shows the outcome of TF-DDRL under various model configurations. Without the use of both PER and Transformer, the native IMPALA requires approximately 90 iterations to converge to the optimal solution discovered in the experiment. The convergence speed slightly improves when only PER is employed. However, with the exclusive employment of the Transformer, TF-DDRL demonstrates a significant acceleration in convergence speed, reaching the experiment's optimal solution in around 50 iterations. When both PER and Transformer are used concurrently, TF-DDRL converges in approximately 40 iterations. This shows that TF-DDRL, compared to native IMPALA, can find a better solution to solve the IoT application scheduling problem more efficiently.
\par
\subsubsection{Cost vs Policy Update Analysis}
\label{cvpua}
\begin{figure*}[h]
\begin{subfigure}{0.24\textwidth}
  \centering
  \includegraphics[width=\linewidth]{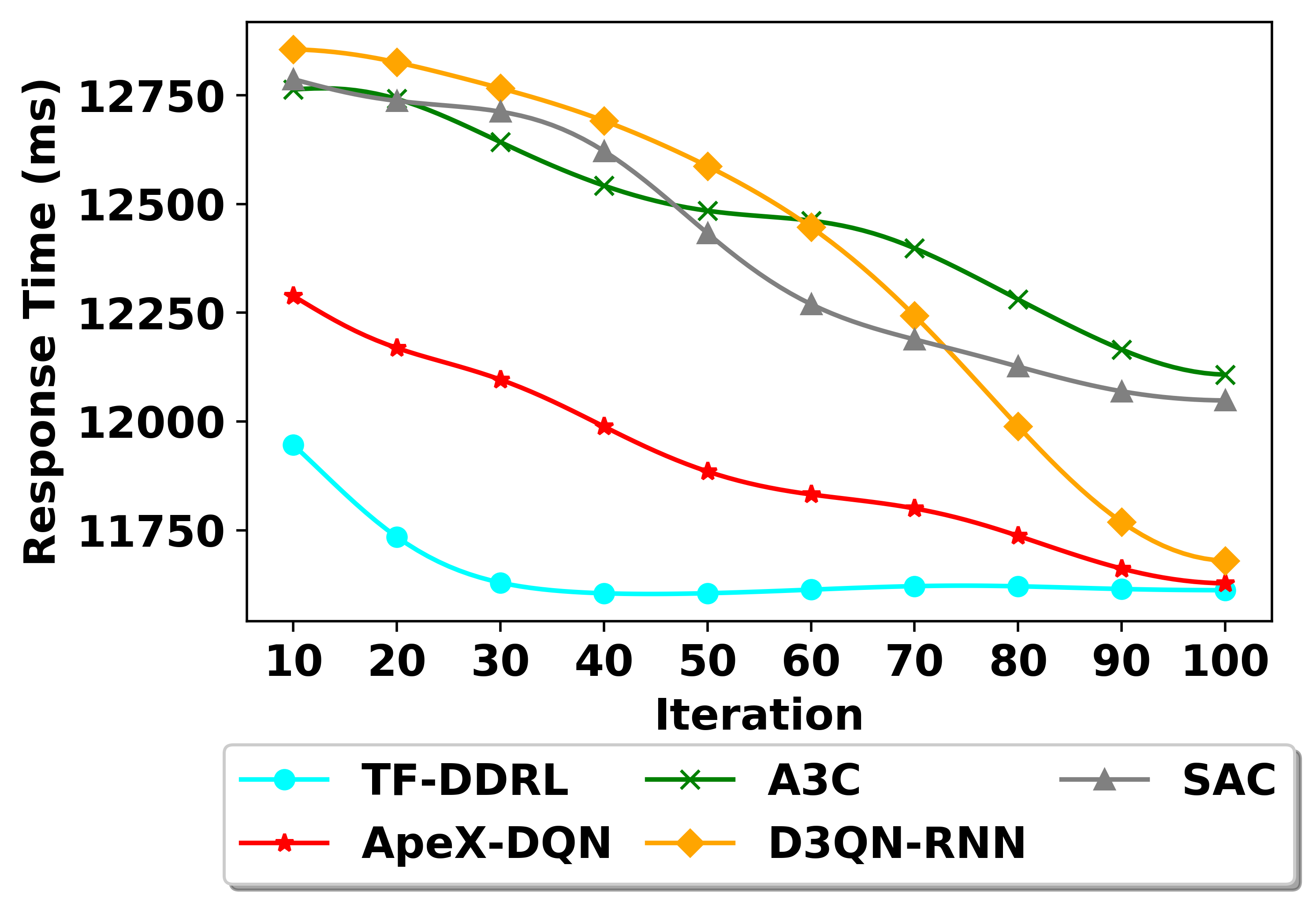}
  \caption{Response time}
  \label{fig:rtt}
\end{subfigure}%
\begin{subfigure}{0.24\textwidth}
  \centering
  \includegraphics[width=\linewidth]{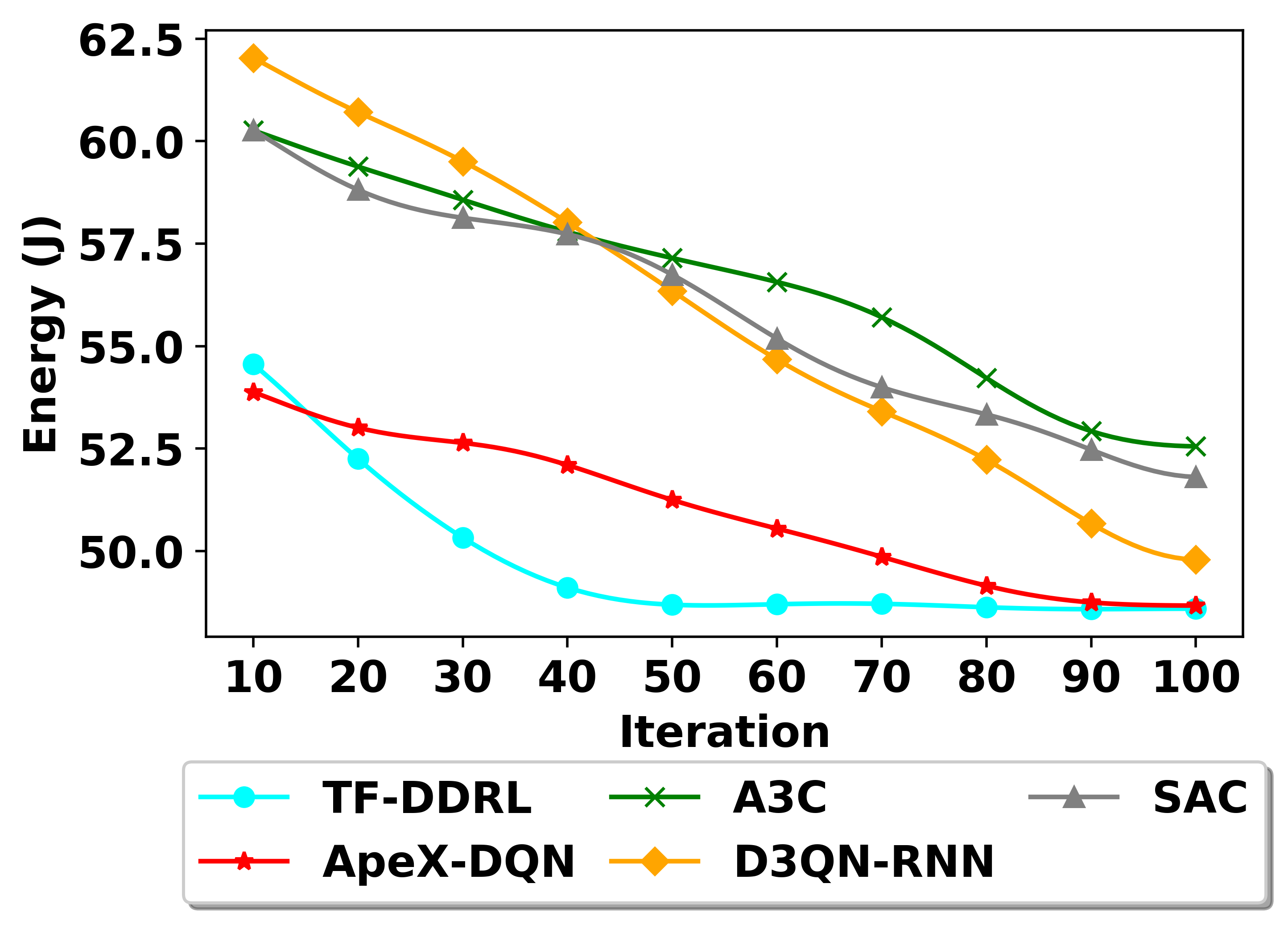}
  \caption{Energy Consumption}
  \label{fig:ent}
\end{subfigure}
\begin{subfigure}{0.24\textwidth}
  \centering
  \includegraphics[width=\linewidth]{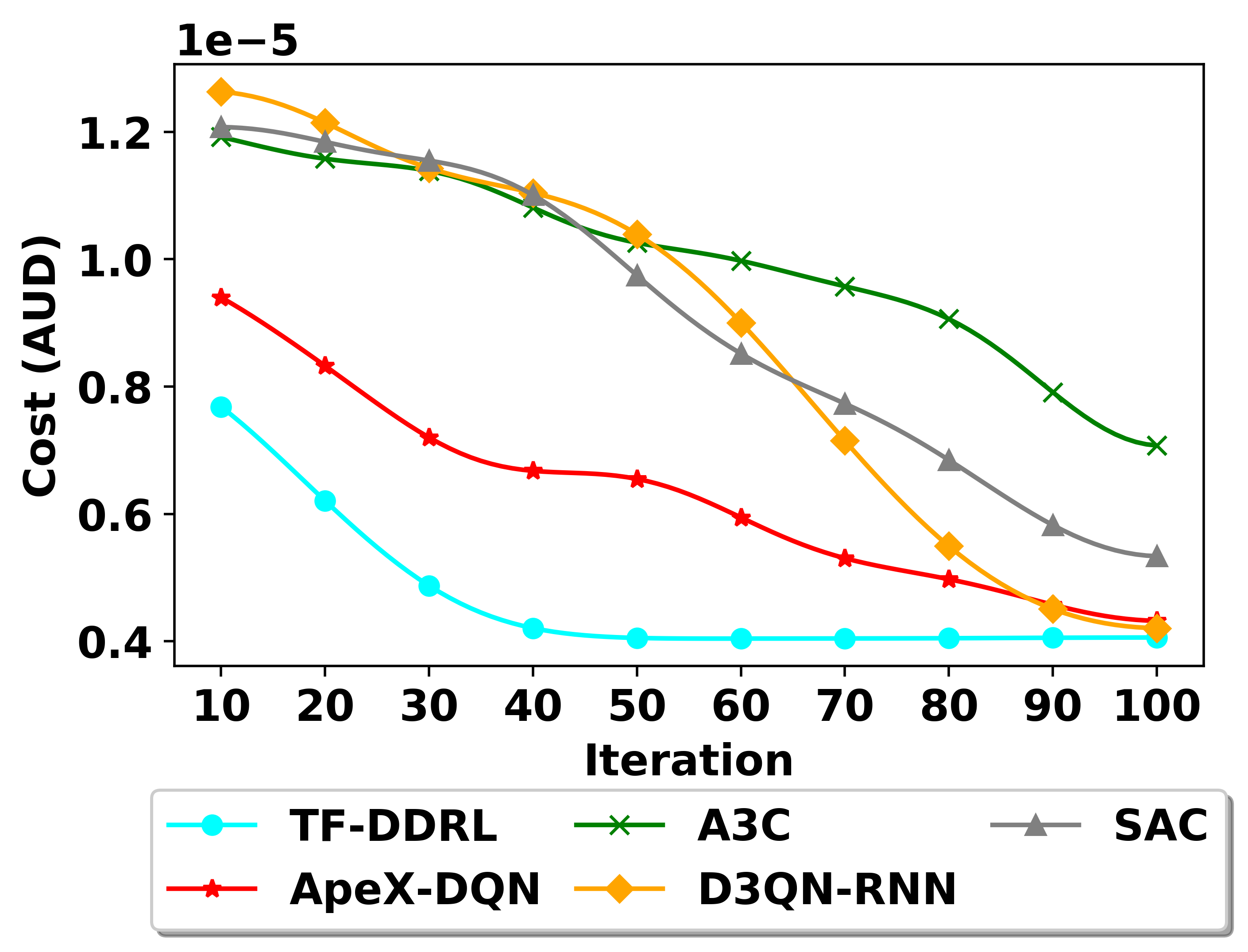}
  \caption{Monetary Cost}
  \label{fig:cot}
\end{subfigure}
\begin{subfigure}{0.24\textwidth}
  \centering
  \includegraphics[width=\linewidth]{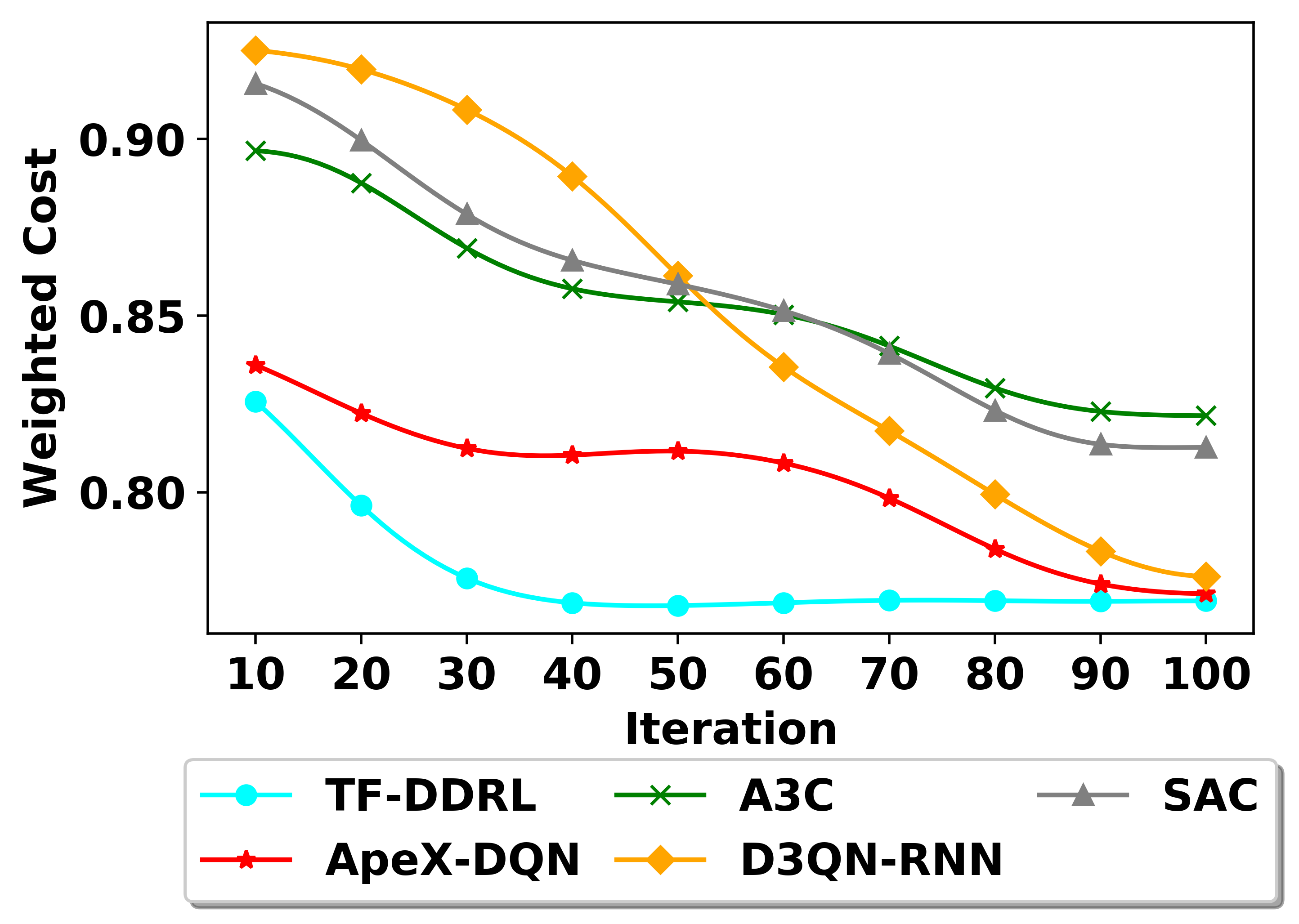}
  \caption{Weighted cost}
  \label{fig:wct}
\end{subfigure}
\caption{Cost vs policy update analysis - training phase}
\label{fig:train}
\end{figure*}

This experiment analyzes the performance of TF-DDRL in various iterations during policy updates. For training purposes, we utilize four applications as detailed in Section \ref{app}, configuring the resolution as 480. The results, showing the policy cost versus updating iteration, are presented in Fig. \ref{fig:train}.
\begin{figure*}[h]
\begin{subfigure}{0.24\textwidth}
  \centering
  \includegraphics[width=\linewidth]{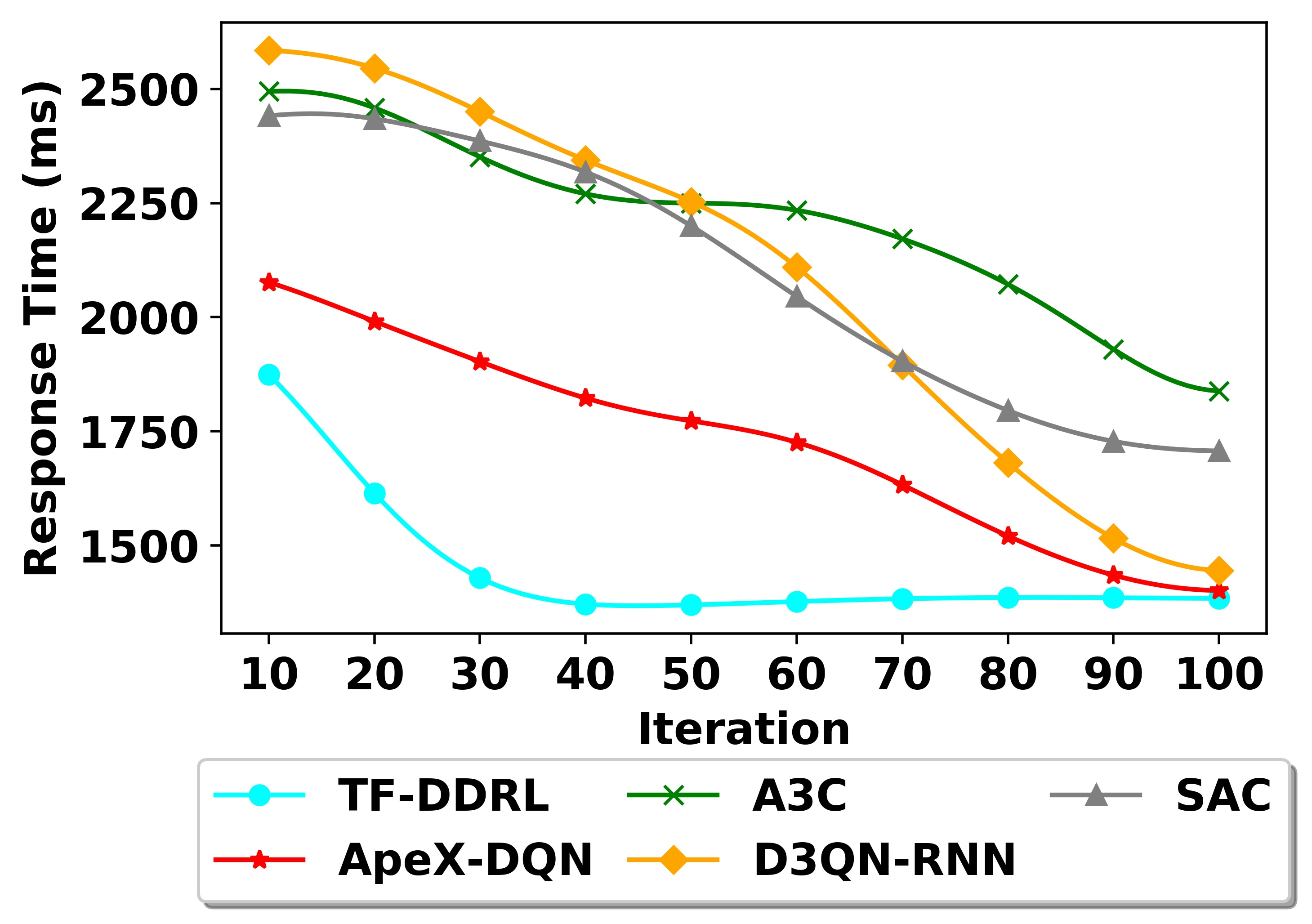}
  \caption{Response time}
  \label{fig:rte}
\end{subfigure}%
\begin{subfigure}{0.24\textwidth}
  \centering
  \includegraphics[width=\linewidth]{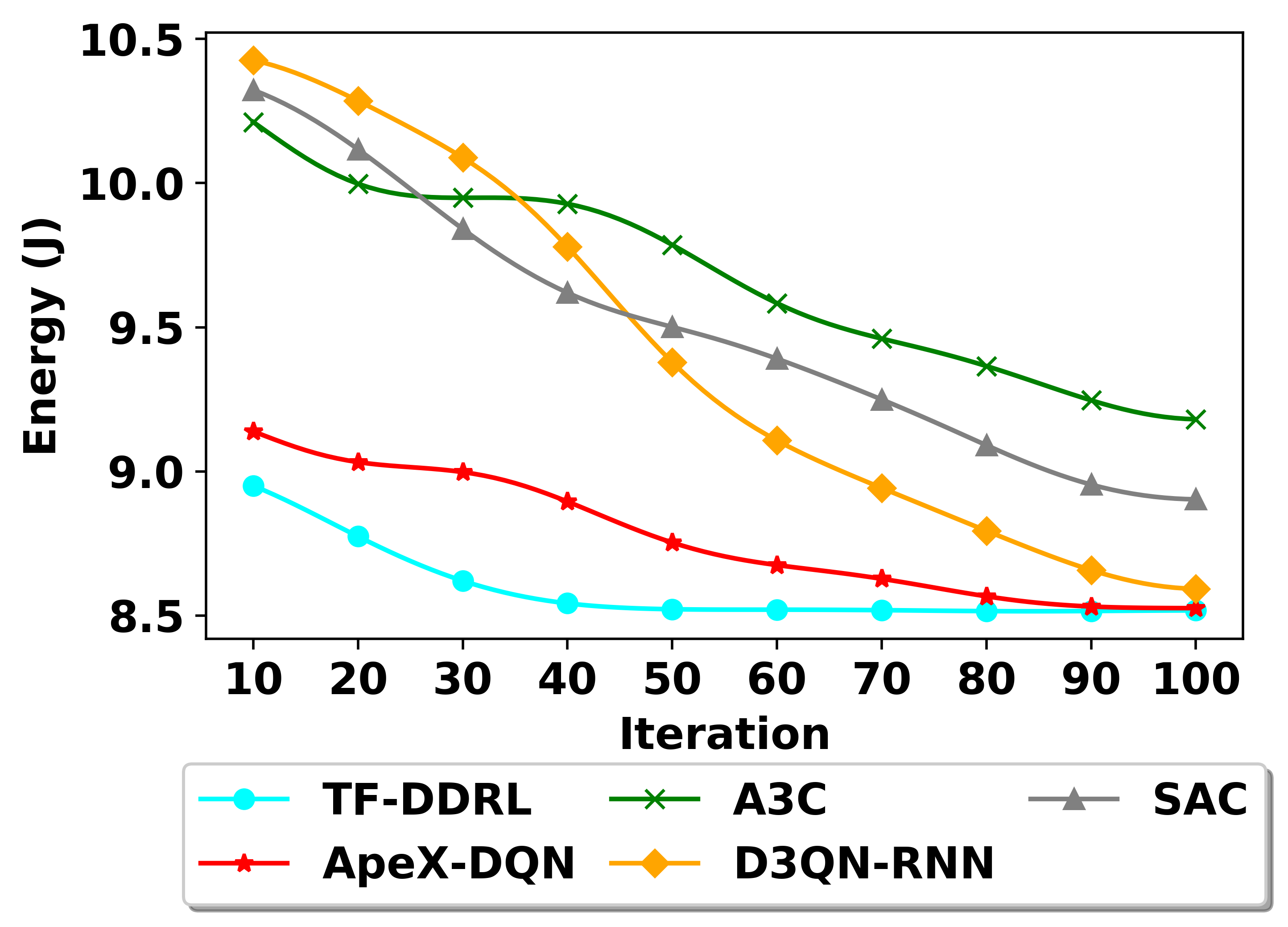}
  \caption{Energy Consumption}
  \label{fig:ene}
\end{subfigure}
\begin{subfigure}{0.24\textwidth}
  \centering
  \includegraphics[width=\linewidth]{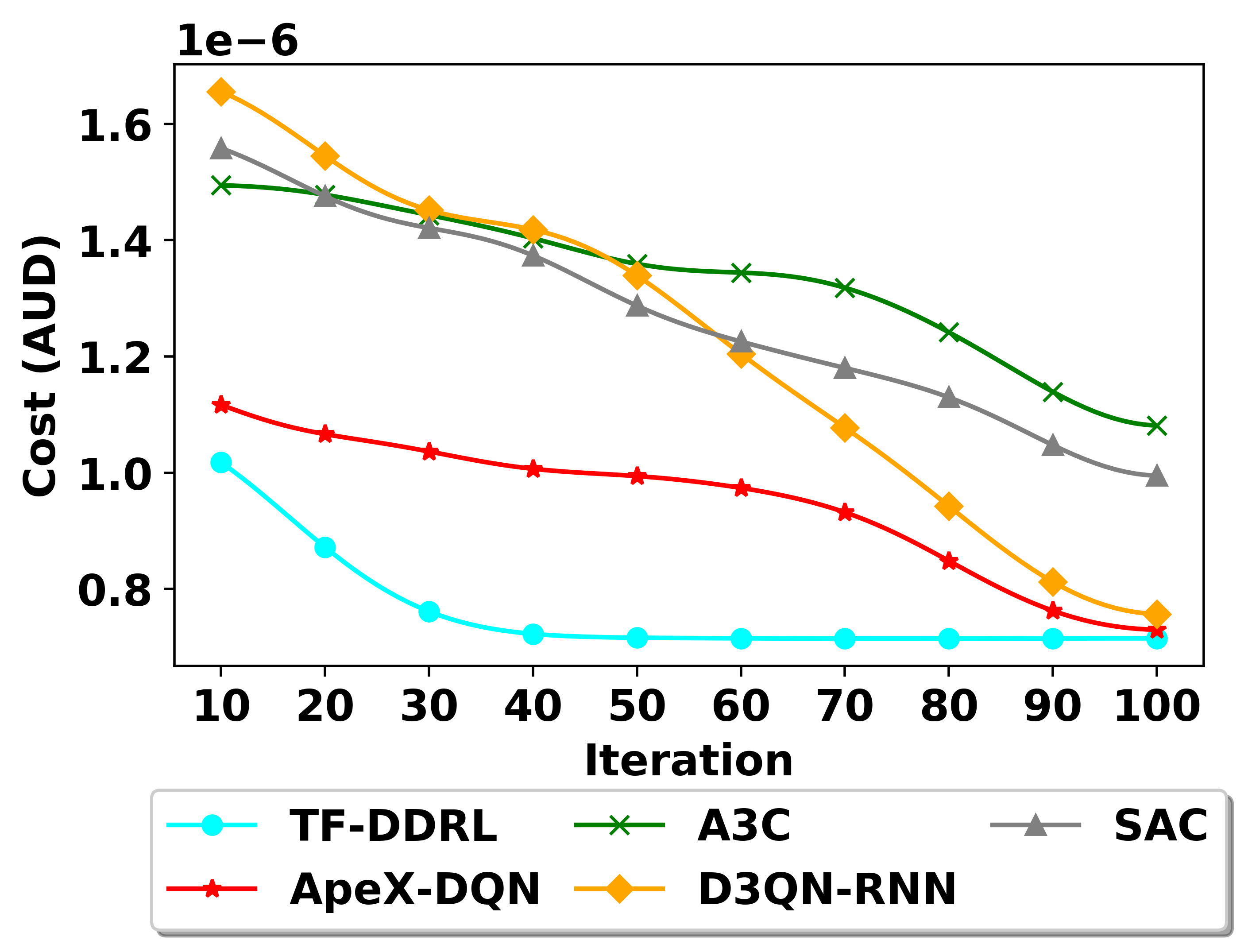}
  \caption{Monetary Cost}
  \label{fig:coe}
\end{subfigure}
\begin{subfigure}{0.24\textwidth}
  \centering
  \includegraphics[width=\linewidth]{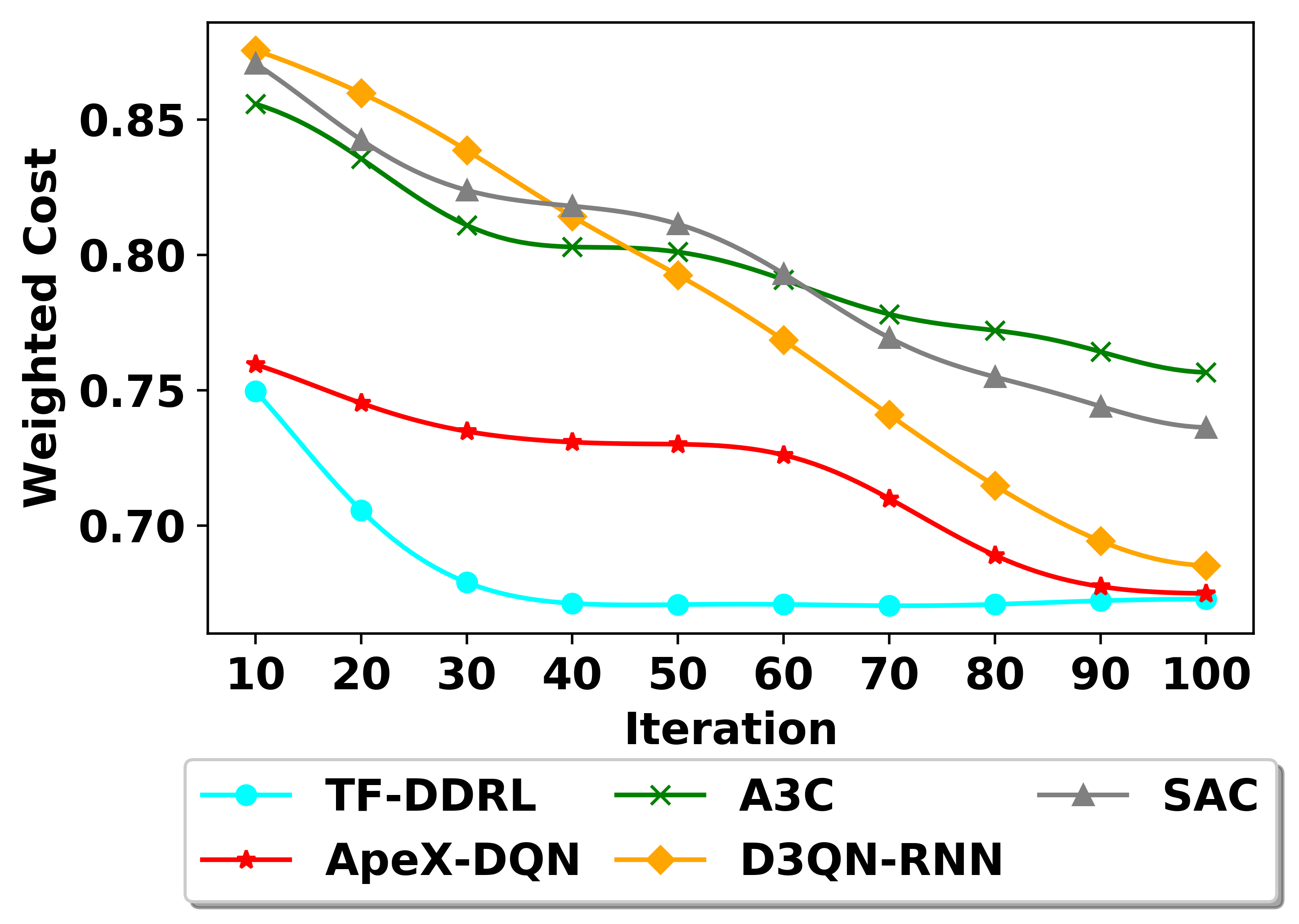}
  \caption{Weighted cost}
  \label{fig:wce}
\end{subfigure}
\caption{Cost vs policy update analysis - evaluation phase}
\label{fig:eva}
\end{figure*}
\par
As shown in Fig. \ref{fig:train}, the optimization costs of all techniques decrease with the increasing number of iterations in different scenarios. However, under different optimization objectives, TF-DDRL shows a faster convergence compared to other techniques. It converges to the best scheduling solution discovered during training in approximately 40 iterations. ApeX-DQN exhibits a slower convergence speed than TF-DDRL but eventually converges to the optimal scheduling solution in 90 iterations. D3QN-RNN converges to the best scheduling solution under monetary cost optimization (Fig. \ref{fig:cot}). Although the costs of A3C and SAC decrease continuously during training, neither of them converges to the optimal solution within 100 iterations.
\par
During evaluation, the resolution is adjusted to 240, altering the IoT application's demands for computing resources compared to the training phase. The results, showing the optimization cost versus the policy update for various algorithms, are presented in Fig. \ref{fig:eva}. It is obvious that similar to the results obtained during the training phase, compared with other techniques, TF-DDRL demonstrates better performance in response time, energy consumption, monetary cost, and weighted cost in the evaluation phase. Also, after 100 iterations of updates for all baseline techniques, none of them can achieve results superior to TF-DDRL. This indicates that TF-DDRL not only converges faster, with significantly less time compared to other techniques but also provides better scheduling results. Except for the ApeX-DQN technique, A3C, D3QN-RNN, and SAC do not converge to the optimal scheduling solution in 100 iterations. Overall, compared to ApeX-DQN results, which is the only baseline technique that converges to the optimal scheduling solution found in the evaluation phase across all optimization objectives, TF-DDRL achieves average performance gains of 60\%, 51\%, 56\%, and 58\% in response time, energy consumption, monetary cost, and weighted cost, respectively.
\par
\subsubsection{Scalability Analysis}
This experiment investigates the impact of different numbers of servers on the scheduling technique for IoT applications. The number of available servers directly impacts the complexity of IoT application scheduling problems, as a higher number of servers leads to a larger action space. To evaluate the scalability performance of TF-DDRL, the experiment uses varying numbers of servers (e.g., 5, 10, 15, 20, 25, 30). Also, other settings are consistent with Section \ref{cvpua}. Due to space constraints and the fact that the results for response time, energy consumption, and monetary cost follow the same patterns as weighted costs, only the results for weighted costs are presented.
\par
\begin{figure*}[!htb]
\minipage[t]{0.24\textwidth}
  \centering
  \includegraphics[width=\linewidth]{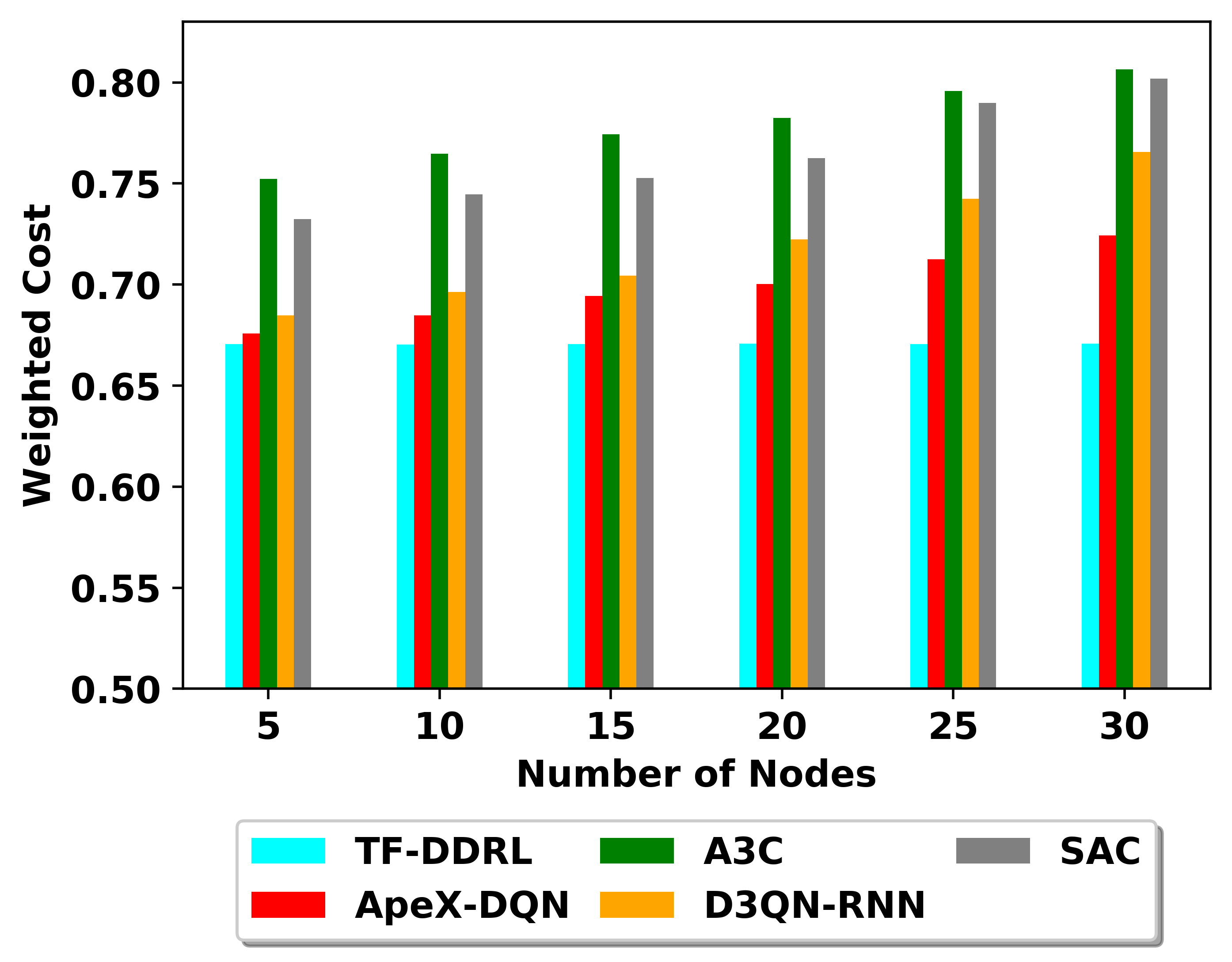}
  \caption{Scalability analysis}\label{fig:sca}
\endminipage\hfill
\minipage[t]{0.24\textwidth}
  \centering
  \includegraphics[width=\linewidth]{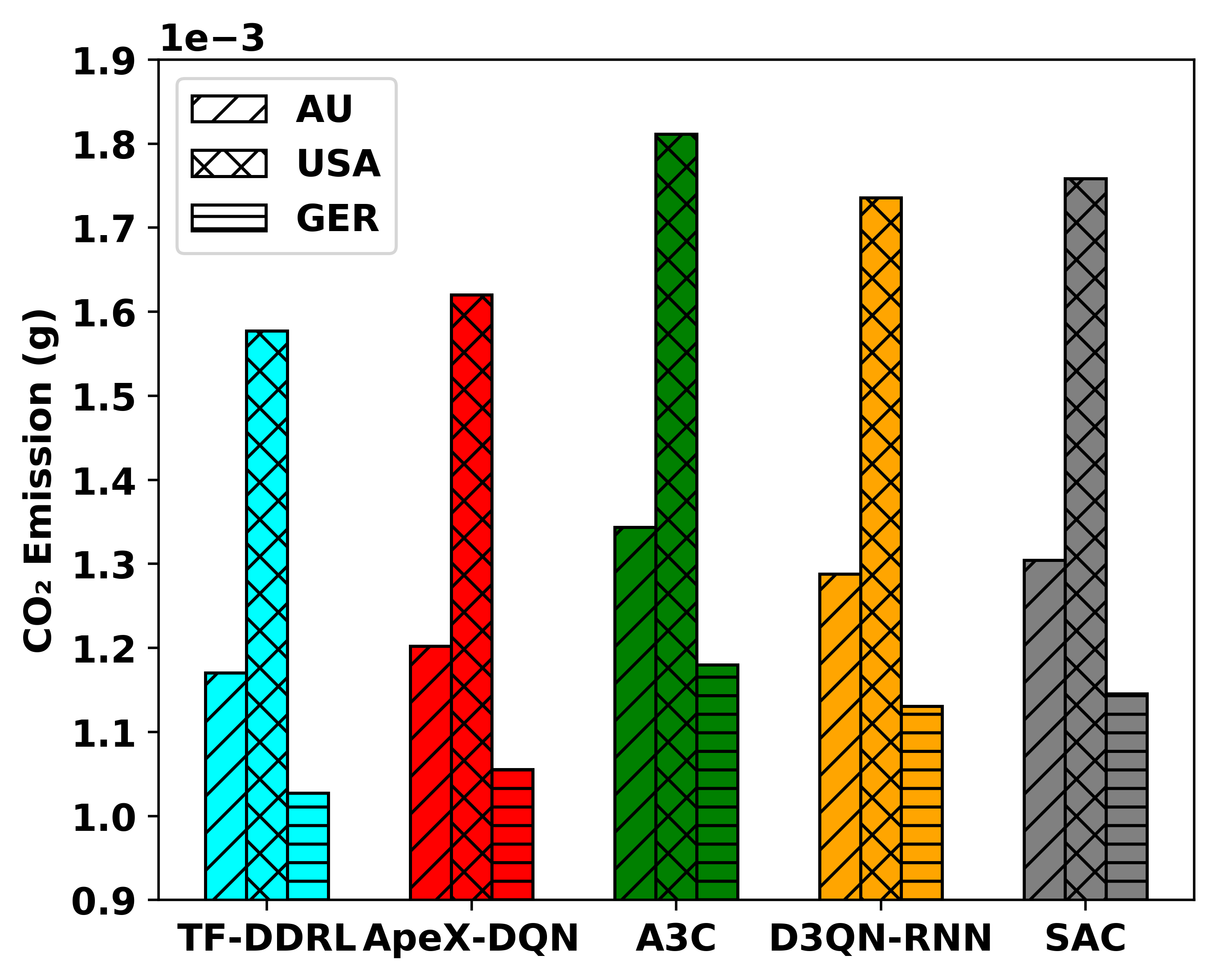}
  \caption{GHE analysis}\label{fig:coc}
\endminipage\hfill
\minipage[t]{0.24\textwidth}
  \centering
  \includegraphics[width=\linewidth]{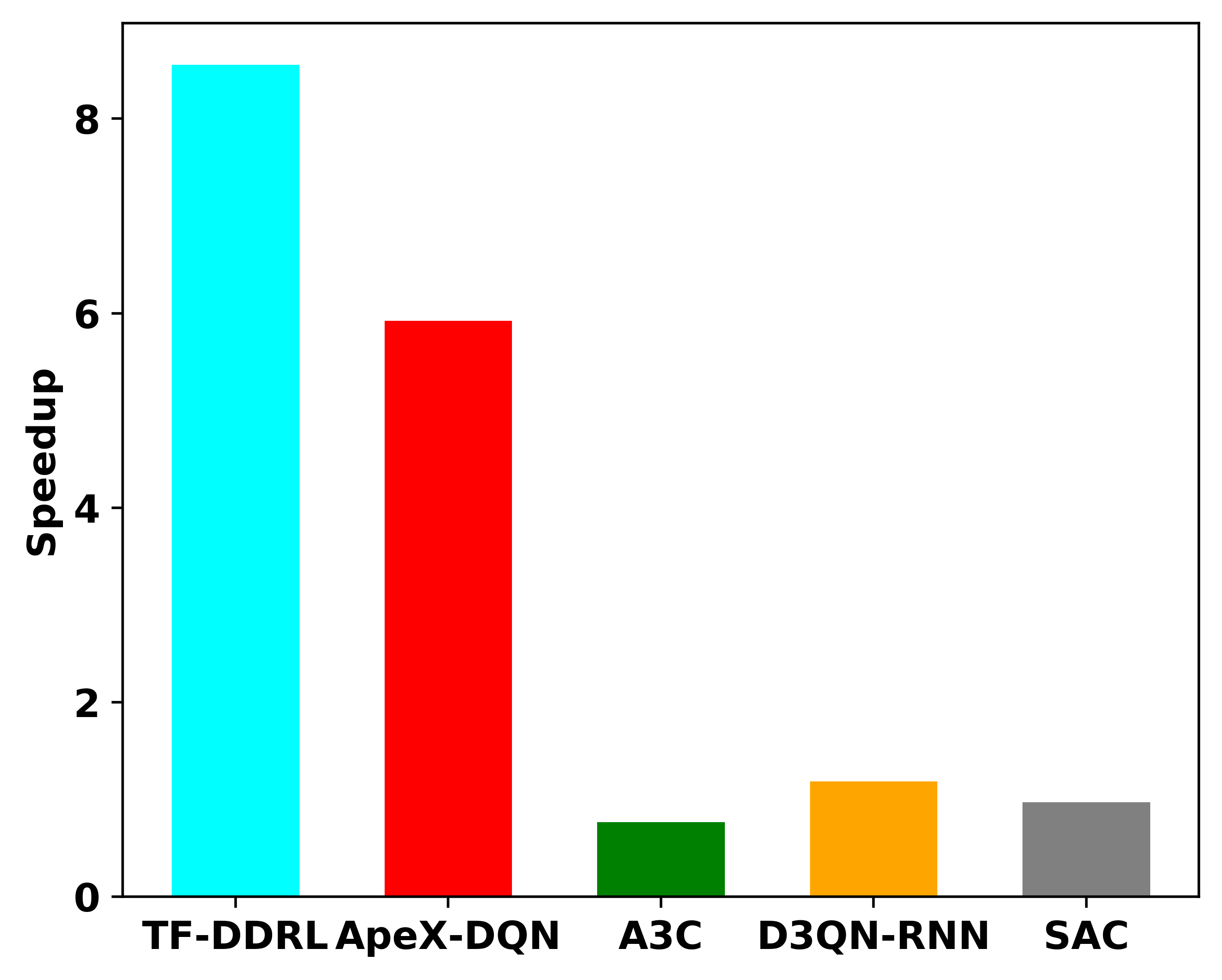}
  \caption{Speedup analysis}\label{fig:spu}
\endminipage\hfill
\minipage[t]{0.24\textwidth}
  \centering
  \includegraphics[width=\linewidth]{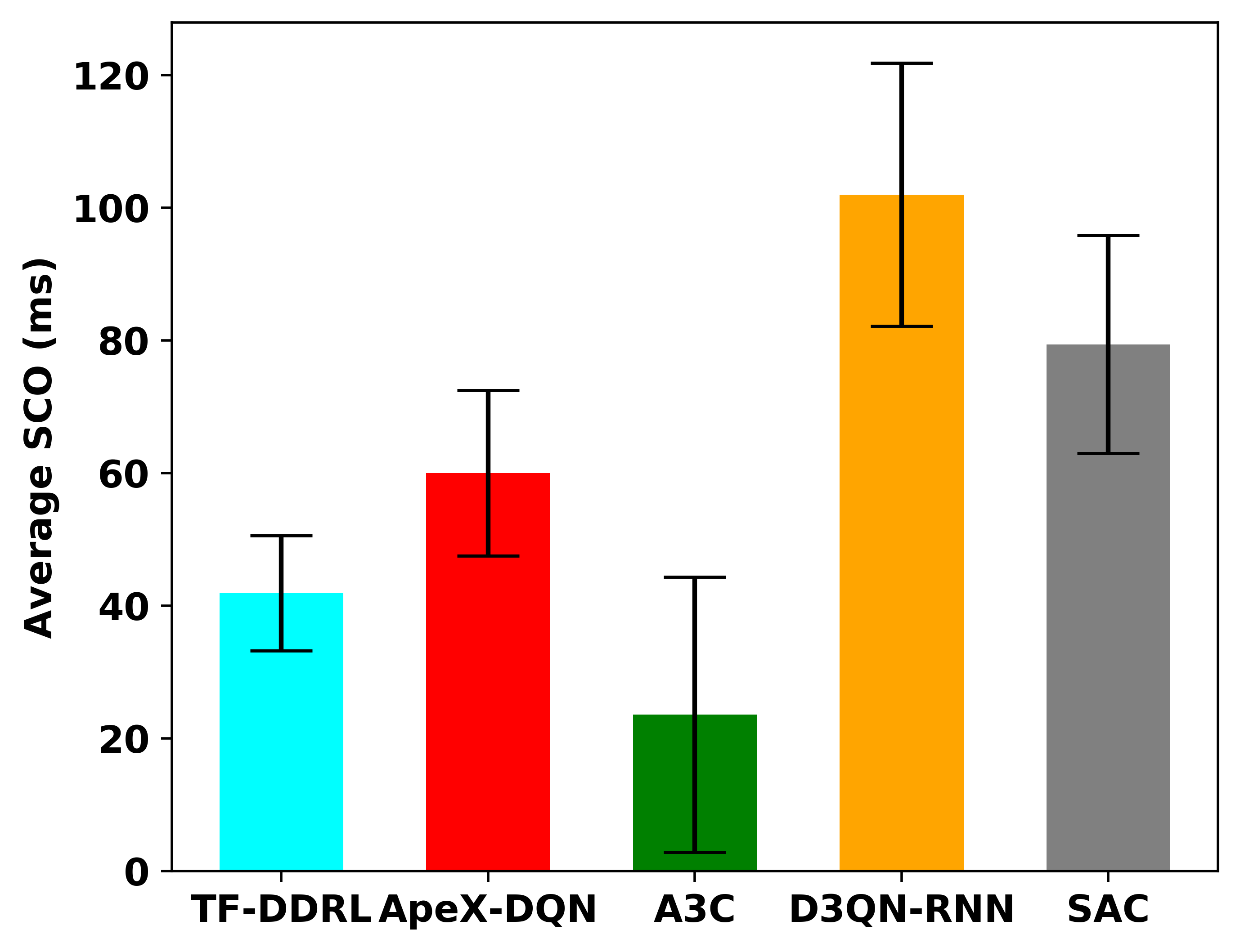}
  \caption{SCO analysis}\label{fig:aso}
\endminipage\hfill
\end{figure*}

Figure \ref{fig:sca} shows the weighted cost optimization results obtained by various techniques after 100 iterations, considering the growth of candidate servers. As the number of servers increases, TF-DDRL consistently outperforms other techniques, converging more rapidly towards superior solutions. This shows that as the system scales up, TF-DDRL demonstrates superior scalability, enabling it to make more effective application scheduling decisions in fewer iteration cycles. In the baseline techniques, ApeX-DQN outperforms other techniques, although weighted costs eventually continue to increase with the growth of available servers.
\par
\subsubsection{Greenhouse Gas Emission Analysis}
This experiment examines the impact of various scheduling techniques based on Greenhouse Gas Emission (GHE). We specifically analyze electricity generation patterns in Australia\footnote{https://www.energy.gov.au/data/electricity-generation}, the US\footnote{https://www.eia.gov/tools/faqs}, and Germany\footnote{https://www.umweltbundesamt.de/themen/co2-emissionen-pro-kilowattstunde-strom-stiegen-in}, considering the associated GHE of various sources involved in electricity production\footnote{https://world-nuclear.org/information-library/energy-and-the-environment/carbon-dioxide-emissions-from-electricity.aspx}. The total GHE is defined as the sum of GHE from the production of electricity from each source \cite{scarlat2022quantification}, shown below:
\begin{equation}
GHE = EC*\sum_{i \in Sources}(U_i*P_i),
\end{equation}
where $EC$ represents the total electricity consumed, $U_i$ represents the amount of greenhouse gas emitted per unit of electricity produced using source $i$, and $P_i$ represents the proportion of the source $i$ in producing electricity.
\par
Figure \ref{fig:coc} presents the GHE associated with different scheduling techniques based on electricity generation in different countries. Notably, TF-DDRL exhibits the lowest GHE, while A3C has more GHE compared to other techniques. Also, the GHE based on the US power generation pattern is substantially higher than that of Australia and Germany. This discrepancy is due to the prevalence of fossil sources, including coal and natural gas, in the US electricity generation pattern. The experiment results show that TF-DDRL can effectively reduce GHE, which can contribute to collective efforts to address climate change, alleviate the impacts of global warming, and foster a healthier and more sustainable natural environment.
\par
\subsubsection{Speedup Analysis}
With the similar experimental configuration outlined in Section \ref{cvpua}, we explore the speedup performance of various techniques. We define the reference time, denoted as $Time_r$, as the time required for the weighted cost of TF-DDRL with an Actor to reach a value of $0.76$. Designating $0.76$ as the reference weighted cost is motivated by the fact that this particular value serves as the smallest weighted cost that can be obtained by all baseline techniques. Additionally, we define $Time_t$ as the time required by each technique to attain the reference weighted cost. So, the speedup $SPU$ for each technique is defined as follows:
\begin{equation}
SPU= \frac{Time_r}{Time_t}.
\end{equation}
\par
The speedup results for all techniques are illustrated in Fig. \ref{fig:spu}. The results demonstrate that TF-DDRL outperforms A3C, D3QN-RNN, and SAC by 7 to 11 times, and it is over 40\% faster than ApeX-DQN. Consequently, TF-DDRL boasts a significantly shorter learning time compared to alternative techniques. This reduction in learning time not only accelerates the convergence of DRL scheduling agents but also enhances adaptability to random and dynamic edge and cloud computing environments.
\par
\subsubsection{Scheduling Overhead (SCO) Analysis}
This experiment investigates the SCO of each technique. We use the same environment settings in Section \ref{cvpua}. For each technique, we run 100 iterations, each containing four IoT applications. Also, the average SCO is defined as $Time_{a} = \frac{Time_{o}}{100}$, where $Time_{o}$ denotes the total overhead of the technique to schedule the IoT applications.
\par
Figure \ref{fig:aso} illustrates the $Time_{a}$ within the 95\% confidence interval for various techniques during the scheduling of IoT applications. The scheduling overhead of TF-DDRL is lower compared to ApeX-DQN, D3QN-RNN, and SAC, but higher than A3C. However, TF-DDRL can converge to a superior solution more quickly with an acceptable overhead compared to A3C. Thus, in heterogeneous edge and cloud computing environments, TF-DDRL proves to be more efficient in scheduling IoT applications. 
\par
\section{Conclusions and Future Work}
\label{conclusions}
In this paper, we proposed a distributed DRL technique, named TF-DDRL, designed to solve DAG-based IoT application scheduling in highly heterogeneous and dynamic edge and cloud computing environments. We formulated the IoT application scheduling problem as an optimization problem and then transformed it into an MDP model, aiming to minimize response time, energy consumption, monetary cost, and weighted cost. We proposed the TF-DDRL, which follows Actor-Critic architecture, incorporating PER and Transformer techniques to decrease exploration costs and enhance convergence speed. TF-DDRL allows multiple parallel and scalable Actors to work simultaneously and share experience trajectories with the Learner, enabling more effective and efficient learning. Also, we used the V-trace off-policy correction method to solve discrepancies between Learner and Actor policies. As demonstrated by extensive experiments, in highly stochastic and heterogeneous computing environments, TF-DDRL possesses better scalability and adaptability, compared to its counterparts. 
The results indicate that TF-DDRL outperforms other DRL-based approaches, demonstrating performance improvements of up to 60\%, 51\%, 56\%, and 58\% in terms of response time, energy consumption, monetary cost, and weighted cost, respectively.
\par
As part of future work, we will consider more aspects of the optimization problem, including system load balancing. Also, we plan to develop a resource management framework based on TF-DDRL for edge and cloud environments, allowing users to customize and evaluate applications and scheduling policies in dynamically heterogeneous environments.
\par


%





\ifCLASSOPTIONcaptionsoff
  \newpage
\fi





\bibliographystyle{IEEEtran}

\vfill


\end{document}